\documentclass[prx,aps,noshowpacs,preprintnumbers,amsmath,amssymb,twocolumn,superscriptaddress]{revtex4-1}

\usepackage{graphicx}
\usepackage{dcolumn}
\usepackage{epsfig}
\usepackage{bbold,bm}
\usepackage{mathtools}
\usepackage{hyperref}

\def\multiset#1#2{\ensuremath{\left(\kern-.3em\left(\genfrac{}{}{0pt}{}{#1}{#2}\right)\kern-.3em\right)}}

\def\bra#1{\mathinner{\langle{#1}|}}
\def\ket#1{\mathinner{|{#1}\rangle}}
\def\inner#1#2{\mathinner{\langle{#1}|{#2}\rangle}}

\def\tr{\mathrm{Tr}}
\def\pf{\mathrm{Pf}}

\def\id{\mathbb{1}} % need bbold,bm package

\def\Texp{\mathrm{T}\hspace{-0.5mm}\exp}

\def\nodag{^{\vphantom{\dag}}}
\def\trans{{\raisebox{-1pt}{{\scriptsize\textsf{T}}}}}

\usepackage{color}

\def\myheading#1{\section{#1}} % PRX
\def\mysubheading#1{\subsection{#1}} % PRX

\makeatother

\begin{document}

%\title{Proximity quenches of topological and gapless wires realize quantum monodromies}
\title{Dynamically Induced Topology and Quantum Monodromies in a Proximity Quenched Gapless Wire}

\author{D. Dahan}
\affiliation{Department of Physics, Ben-Gurion University of the Negev, Beer-Sheva
	8410501, Israel}

\author{E. Grosfeld}
\affiliation{Department of Physics, Ben-Gurion University of the Negev, Beer-Sheva
	8410501, Israel}

\author{B. Seradjeh}
\affiliation{Department of Physics, Indiana University, Bloomington, Indiana 47405,
	USA}
\affiliation{Max Planck Institute for the Physics of Complex Systems, N\"othnitzer Stra\ss e 38, Dresden 01187, Germany}

\begin{abstract}
We study the quench dynamics of a topologically trivial one-dimensional gapless wire following its sudden coupling to topological bound states. We find that as the bound states leak into and propagate through the wire, signatures of their topological nature survive and remain measurable over a long lifetime. Thus, the quench dynamically induces topological properties in the gapless wire. Specifically, we study a gapless wire coupled to fractionally charged solitons
or Majorana fermions and characterize the dynamically induced topology in the wire, in the presence of disorder and short-range interactions, by analytical and numerical calculations of the dynamics of fractional charge, fermion parity, entanglement entropy, and fractional exchange statistics. In a dual effective description, this phenomenon is described by correlators of boundary changing operators, which, remarkably, generate topologically non-trivial monodromies in the gapless wire, both for abelian and non-abelian quantum statistics of the bound states.  
\end{abstract}

\date{\today}

\maketitle

\myheading{Introduction}%
\vspace{-2mm}
Ordered phases of matter typically exhibit a proximity effect: when interfaced with a normal, gapless system, an exponentially thin layer of the normal system is \emph{induced} with the proximate order. This phenomenon is well understood for symmetry-broken phases with a local order parameter \cite{holm1932proximity}, in terms of local interactions induced in the normal system by quantum tunneling over a coherence length of the proximate order. When topological phases, which lack a local order parameter, are interfaced with a normal insulator, the interface hosts topological bound states (TBSs) with unconventional assignments of quantum numbers, such as fractional charge or unpaired helicity \cite{jackiw1976solitons,arovas1984fractional,haldane1988model,kane2005topological,bernevig2006quantum}. If we replace the normal insulator with a normal, gapless conductor, the TBSs lose their integrity in equilibrium and become extended on the normal side. Thus, it would seem that topology cannot be induced by local proximity in space.

This spatial view in equilibrium, however, is blind to the temporal correlations out of equilibrium. We can think of the process of interfacing the topological phase with a gapless system as a rapid quench of the boundary conditions of the system, in time. Once the interface is in place, the TBSs become free to move through the gapless medium \cite{vasseur2014universal,sacramento2014fate,dahan2017non}. Could there be a \emph{temporal} topological proximity effect, whereby proximity to the topological phase induces the normal system with some topological features \emph{dynamically} over some coherence time?

In this work, we answer this question in the affirmative. 
The scaling laws of the short-time dynamics after a local quench were highlighted in \cite{vasseur2014universal} in terms of the conformal dimensions of boundary-changing operators (BCOs) implemented by the quench. Global quenches of topological states were considered in~\cite{degottardi2011topological,rajak2014survival,hegde2015quench}.
What we show here is that the \emph{long-time} dynamics uncovers some of the most interesting features of the correlators of the BCOs, including wavefunction monodromies of both Abelian and non-Abelian nature. These monodromies are encoded in the revivals of the quantum dynamics of the Fermi sea after the quench, with different periods exhibiting unique patterns of charge fluctuations, fermion parity, entanglement, fidelity, and phase discontinuities. We dub this behavior ``{dynamically induced topology},''  and characterize it through the evolution of fractional quantum numbers and entanglement entropy in a gapless one-dimensional wire (GW). We demonstrate that, remarkably, the TBSs form mobile anyons within the GW and manifest their quantum statistics even in this one-dimensional geometry. The anyons maintain their localization during a coherence time which extends over many multiples of the return time of the GW. In this way, we formulate the temporal evolution of the induced topology in the topologically trivial, gapless system.

%, $\tau_r=2\ell/v_F$, where $\ell$ is the length of the wire and $v_F$ its Fermi velocity. 

%We explain this behavior using a low energy effective theory, where the quench is realized by the application of boundary changing operators (BCOs). We show that the observed patterns are manifestations of monodromies between conformal blocks of the correlators of the BCOs. For abelian monodromies, we characterize the associated abelian  phase and show that it encodes the quantum statistics of the TBSs. 

%%%%%%%%%%%%%%%%%%%%%%%%
%%%%%%%%% FIG. 0 %%%%%%%%%%%
%%%%% EFFECTIVE SETUP %%%%%%%%
%%%%%%%%%%%%%%%%%%%%%%%%
\begin{figure}[t]
%	\hspace*{-0.15in}
\begin{center}
	\includegraphics[width=3.4in]{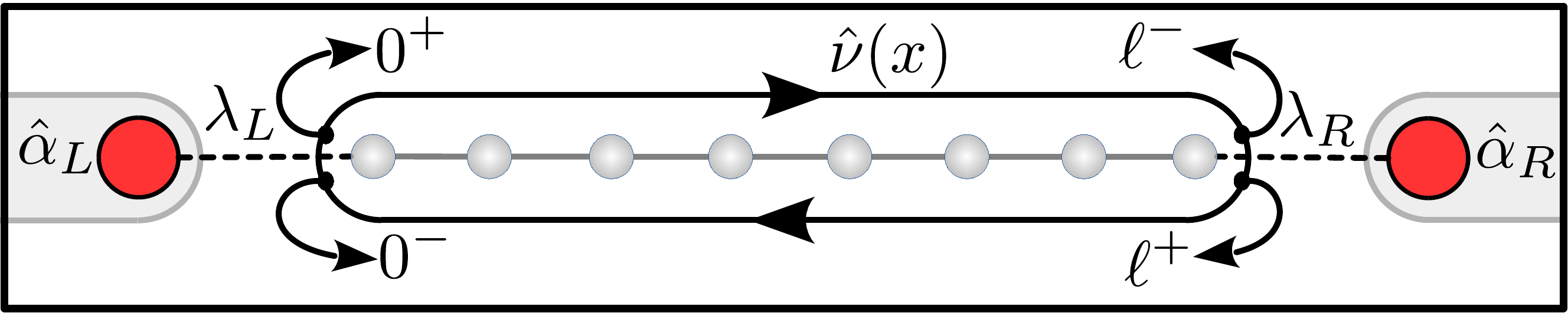}
	\caption{
		%Panel (a), the setup: 
		Setup: a gapless wire of length $\ell$ is coupled to bound states $\hat\alpha_{L,R}$ at its endpoints. The couplings $\lambda_{L,R}$ are switched according to a quench protocol. A chiral field $\hat\nu$ describes the propagating modes in the wire. 
		%Panels (b) and (c) compare exact diagonalization (solid black lines) with effective theory results (dotted orange lines). Panel (b): density at the rightmost site of the left dimerized system ($m_L=0.8w$, $N_L=50$), coupled via $\lambda_L=0.2w\,\theta(t)$ to a GW ($N=300$), with $\lambda_R=0$ for all $t$. Panel (c): fermion parity of the left Kitaev chain, for two Kitaev chains ($N_{L, R}=20$, $\Delta_{L, R}=-0.7w$) coupled via $\lambda_{L, R}=0.2w\,\theta(t)$ to a GW ($N=250)$. 
	} 
	\label{fig:setup} 
	\end{center}
	\vspace{-8mm}
\end{figure}
%%%%%%%%%%%%%%%%%%%%%%%%

\vspace{-4mm}
\myheading{Setup and Model}%
\vspace{-2mm}
Our setup consists of a multipart system composed of a wire that is gapless in the thermodynamic limit and one or two topological wires labelled by $a\in\{L,R\}$ that support TBSs, either fractional or Majorana fermions (see Fig.~\ref{fig:setup}). The couplings $\lambda_a$ between the topological and gapless wires are time-dependent according to a prescribed quench protocol. We consider a \emph{single quench} by which all components of the system are suddenly coupled at time $t_0$ and remain coupled thereafter, as well as a \emph{double quench} by which a single quench is applied and then all original components are disconnected at a later time $t_1$. 

We model the GW as a tight-binding chain (with $N$ sites) of spinless fermions created by $\hat{c}_r^\dag$ at position $r$ with nearest-neighbor tunneling $w$ and a Hubbard interaction $u$. The topological wires are modeled as Su-Schrieffer-Heeger (SSH) chains \cite{su1979solitons,meier2016observation} with nearest-neighbor tunneling $w_a+(-1)^r m_a$ where $m_a$ is the tunneling modulation, supporting fractional-fermion TBSs, or Kitaev (K) chains \cite{kitaev2001unpaired,lieb1961two} with nearest-neighbor tunneling $w_a$ and pairing $\Delta_a$, supporting Majorana-fermion TBSs. 

The models are solved analytically using effective low-energy and conformal field theory as well as numerically by exact diagonalization for the non-interacting GW, including disorder (see Appendix~\ref{app:exact-time-evolution} for details), and using time-dependent density matrix renormalization group (tDMRG) for the interacting clean GW. 

\myheading{Exchange Statistics}%
TBSs can have fractional exchange statistics in two spatial dimensions when adiabatically braided~\cite{read2000paired,ivanov2001non,kitaev2003fault,alicea2011non}. Remarkably, we discover signatures of the exchange statistics in our one-dimensional, non-equilibrium setup. For the case of two Majorana fermions with no phase difference, the non-Abelian statistics satisfied by the Majorana modes is manifested as a doubling of the expected period for revivals of the fidelity~\cite{dahan2017non}. The case of the fractional solitons of the SSH chain is more subtle, as the information about the exchange statistics is contained only in the accumulated phase. 

 To reveal this \emph{non-adiabatic} exchange statistics, we consider two topological wires coupled via a single quench to opposite ends of a GW and examine the resulting fidelity as well as the phase accumulated in consecutive revivals. Denoting the state of the system by $|\Omega_{p_Lp_R}(t)\rangle$, where $p_L,p_R\in\{0,1\}$ are the initial fermion parities of, respectively, the two fractional solitons and the two superconductors for the SSH and Kitaev chains, we define the overlap of the evolved state with the initial state by $\mathcal{O}_{q_Lq_R,p_Lp_R}(t)=\langle\Omega_{q_Lq_R}(t)|\Omega_{p_Lp_R}(0)\rangle$~\cite{onishi1966generator,robledo2009sign} and extract the probability $|\mathcal{O}_{q_Lq_R;p_Lp_R}(t)|^2$ as well as the relative phase between two processes with different initial and final parities~\cite{bolukbasi2012rigorous}
\begin{eqnarray} \label{eq:stat-phase}
\theta^{q_Lq_R,p_Lp_R}_{q'_Lq'_R,p'_Lp'_R}(t)=\arg\frac{\mathcal{O}_{q_Lq_R,p_Lp_R}(t)}{\mathcal{O}_{q'_Lq'_R,p'_Lp'_R}(t)},
\end{eqnarray}
which encodes various non-Abelian statistical phases by removing the dynamical phase. 

For example, for the SSH solitons, the relative phase between processes with $p_L=p_R=1=q_L=q_R$ and $p'_L=1=q'_L$, $p'_R=0=q'_R$ encodes the Abelian statistical phase accumulated by taking a soliton around another (of the same fractional charge),
%%%% -- PRL $
\begin{equation}
\theta_s^\text{SSH} \equiv \frac12\theta_{10,10}^{11,11}.
\end{equation}
%%%% -- PRL $
For the Majorana fermions, starting with $p_L=p_R=p'_L=p'_R=0$, one can acquire a superposition of $q_L=q_R=1$ and $q'_L=q'_R=0$ final parities with the non-Abelian statistical phase
%%%% -- PRL $
\begin{equation}
\theta_s^\text{K} \equiv \theta_{00,00}^{11,00}.
\end{equation}
%%%% -- PRL $

%%%%%%%%%%%%%%%%%%%%%%%%
%%%%%%%%% FIG. 4 %%%%%%%%%%%
%%%%%% FIDELITY & PHASE %%%%%%%
%%%%%%%%%%%%%%%%%%%%%%%%
\begin{figure}
\begin{center}
	\includegraphics[width=3.4in]{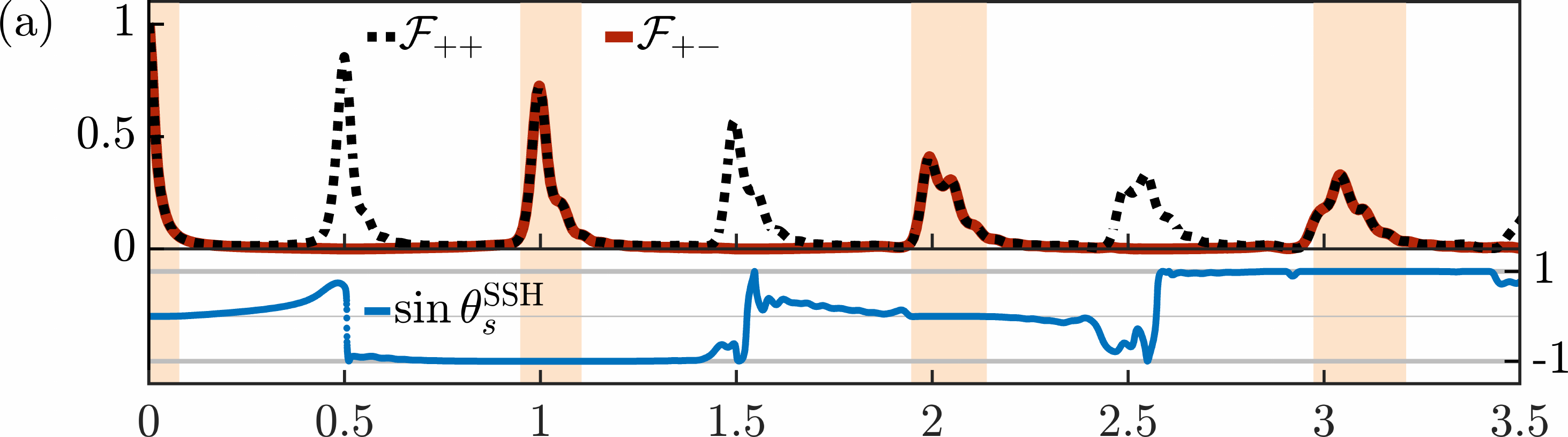}\\
	\vspace*{0.075cm}
	\includegraphics[width=3.4in]{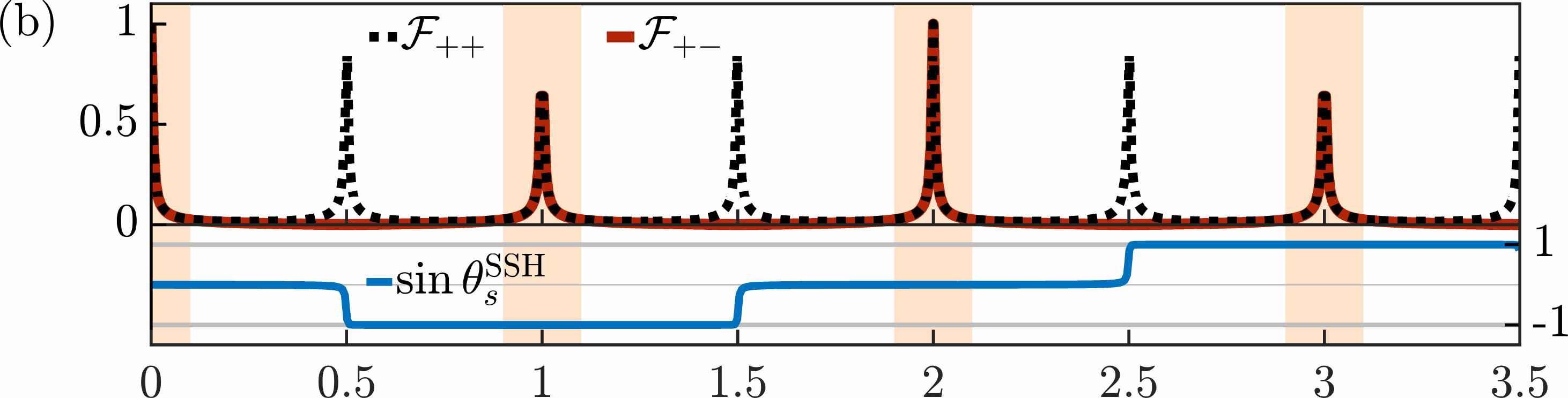}\\
	\vspace*{0.075cm}
	\includegraphics[width=3.4in]{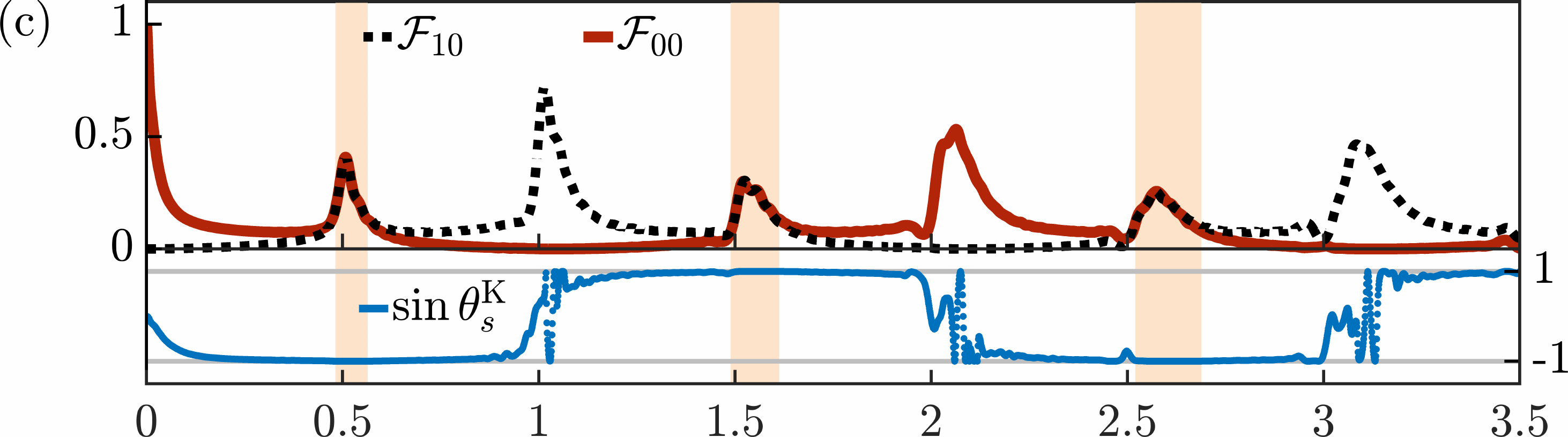}\\
	\vspace*{0.075cm}
	\includegraphics[width=3.4in]{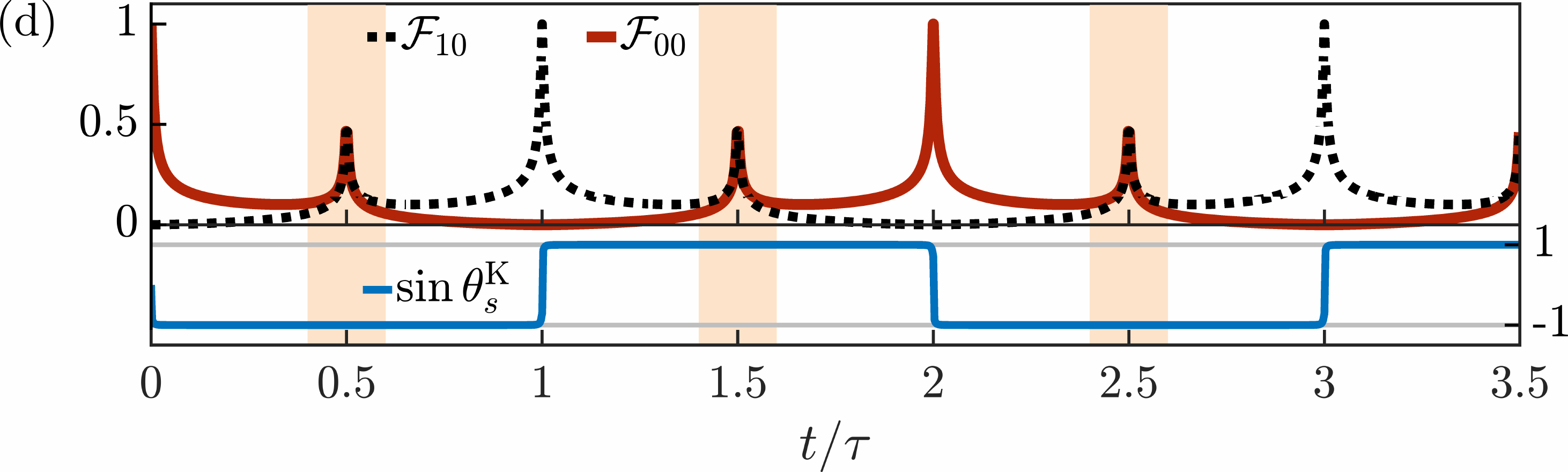}
\vspace{-5mm}
	\caption{
		Fidelity and statistical phase. 
		Panels (a) and (b) show the fidelity $\mathcal{F}_{++}=|\mathcal{O}_{11,11}|^2$ (dashed black), the fidelity $\mathcal{F}_{+-}=|\mathcal{O}_{10,10}|^2$ (red) and the statistical phase $\theta_s^\text{SSH}=\frac{1}{2}\theta^{11,11}_{10,10}$ (%divided by $\pi$,
		blue)
		in a GW ($N=100$) after a single quench with two SSH chains  ($N_{L, R}=50$, $m_{L, R}=0.8w$, $\lambda_{L, R}=0.5w$).
		Panels (c) and (d) show the fidelity $\mathcal{F}_{00}=|\mathcal{O}_{00,00}|^2$ (dashed black), the parity switching probability $\mathcal{F}_{10}=|\mathcal{O}_{11,00}|^2$ (red) and the statistical phase $\theta_s^\text{K}=\theta^{11,00}_{00,00}$ (%divided by $\pi$,
		blue) in a GW ($N=200$) after a single quench at $t_0=0$ with two Kitaev chains ($N_{L, R}=50$, $\Delta_{L, R}=-0.3w$, $\lambda_{L, R}=w$).
		Panels (a) and (c) show the numerical results of exact diagonalization and panels (b) and (d) show the analytical result obtained from Eq.~\eqref{eq:fid-bco}.
	}      
	\label{fig:FidelityAndPhase} 
\end{center}
\vspace{-8mm}
\end{figure}
%%%%%%%%%%%%%%%%%%%%%%%%

We obtain overlaps and phases numerically, as detailed in Appendix~\ref{app:overlap-fidelity-phase}. For both the Kitaev and the SSH chains we find non-trivial statistical phases, see Fig.~\ref{fig:FidelityAndPhase}. The effect requires no coupling between TBSs and the mechanism is elucidated by the low energy effective theory developed below: following the quench, the two TBSs leak into the GW and propagate through opposite chiral channels, exchanging chiralities from one traversal to the next. Through this process they effectively braid, keeping their chirality intact and hence revealing their exchange statistics. The TBS dissipation occurs mainly through amplitude decay into dispersive modes of the GW, while their phase remains coherent over many revivals.

\vspace{-3mm}
\myheading{Effective Theory and Quantum Monodromies}%
We now turn to describe the effective theory and derive the dynamics of the fields and the TBSs. In the following we work with the natural units $\hbar=e=1$. The non-interacting GW, $u=0$, is described by the linearized Hamiltonian 
%%%% -- PRL $
\begin{equation}
\hat H_{\text{GW}}^{\text{eff}}=-iv_F\int_{-\ell}^\ell dx\, \hat\nu^\dagger(x)\partial_x\hat \nu(x),
\end{equation}
%%%% -- PRL $
written in terms of a chiral field $\hat\nu$, with $\{\hat\nu(x),\hat\nu^\dagger(x') \} = \delta(x-x')$ and the boundary condition $\hat\nu(\ell) = \zeta \hat\nu(-\ell)$, $\zeta = e^{2ik_F \ell} = \pm 1$ at half-filling. When working with fractional solitons (SSH) or Majorana fermions (Kitaev) TBSs, we choose to take the chiral field as fermionic $\hat\nu=\hat\psi$ (SSH) or Majorana $\hat\nu=\frac1{\sqrt{2}}\hat\eta=\hat\nu^\dagger$ (Kitaev). This field couples to the TBSs via
%%%% -- PRL $
\begin{equation}
\hat H_{\text{t}}^{\text{eff}}= 2 \sum_{a\in\{L,R\}}\lambda_a(t) \hat\nu(x_a) \hat \alpha_a^\dagger+\text{h.c.},
\end{equation}
%%%% -- PRL $
where $\hat \alpha_a$ denotes a fermion annihilation operator $\hat f_a$ (SSH) or a Majorana zero mode $i  \hat\gamma_a/\sqrt{2}$ (Kitaev) with $\hat\gamma_a^2=\frac12$ interfaced at $x_{L}=0$ and $x_{R}=\ell$ to the GW. See Appendix~\ref{app:effective-theory} for details and derivations. 
%??? check EOM, a factor of sqrt{2} might be handy for the H_t eq. but then the anti-commutation relation doesn't hold for EOM. I suspect that something is not well tunned with these definitons. 
%Here $a=L$ ($R$) denotes the TBS interfaced with the GW to the left (right) at position $x_{L}=0$ ($x_{R}=\ell$). 

Before the quench, $t<t_0=0$, $\hat \alpha_a(t)=\hat \alpha_a(0)$ is constant, and $\hat\nu$ is the free field $\sum_\omega e^{i(kx-\omega t)}\hat \nu_0(\omega)$, where $\hat \nu_0(\omega)$ are the modes of the unperturbed GW with energy $\omega$, and, in the linearized model, $k=\omega/v_F$. The equations of motion for $t>0$ %prescribe
%%%%%
\begin{align}
\nonumber 
&\partial_t\hat \nu(x,t) = -v_F\partial_x\hat \nu(x,t)+ 2i\sum_a \lambda_a\hat \alpha_a(t)\delta(x-x_a),\\ \label{eq:motion}
& \partial_t\hat \alpha_a(t)=2i\lambda_a\hat \nu(x_a,t)=i\lambda_a[\hat \nu(x_a^+,t)+\hat \nu(x_a^-,t)].
\end{align}
%%%%%
Since the GW modes all propagate at $v_F$, we can model the scattering off the TBS as a time-periodic perturbation occurring at regular intervals of order of return time $\tau_r = 2\ell/v_F$. Complemented by the boundary condition $\hat \nu(x+v_F\tau_r,t)=\zeta\hat \nu(x,t)$, we can obtain the full solution.

Before we present the solutions to Eq.~\eqref{eq:motion}, we note that in the low energy limit $\lambda_a \gg v_F/\sqrt{\ell}$ we must have $\hat\nu(x_a^-)=-\hat\nu(x_a^+)$. That is, the quench introduces a $\mathbb{Z}_2$ branch cut into the field. The non-interacting GW is described 
by two copies of the chiral part of the $\mathbb{Z}_{2}$ Ising conformal
field theory~\cite{francesco2012conformal}. Each copy contains three primary fields, the identity
$I$ (with conformal dimension $h_{I}=0$), a Majorana fermion field $\eta(x,t)$
($h_{\eta}=1/2$), and the twist field $\sigma(x,t)$ ($h_{\sigma}=1/16$), which acts as the BCO generating the quench in this limit. Therefore, starting in $|\Omega_0\rangle$,
the state of the GW following the quench with two TBSs is $|\Omega(t)\rangle=\sigma(\ell,t)\sigma(0,t)|\Omega_0\rangle$. Therefore, the overlap (including the phase),
%%%%%%
\begin{equation}\label{eq:fid-bco}
	\langle\Omega(t)|\Omega(0)\rangle=\langle\Omega_0|\sigma(0,t)\sigma(\ell,t)\sigma(\ell,0)\sigma(0,0)|\Omega_0\rangle,
\end{equation}
%%%%%%
is given by correlators of the twist operator and realizes quantum monodromies. We present explicit expressions for this overlap in Appendix~\ref{app:effective-theory}. The results are plotted in Fig.~\ref{fig:FidelityAndPhase}\,, showing remarkable agreement with the numerical results.

%%%%%%%%%%%%%%%%%%%%%%%%
%%%%%%%%% FIG. 3X %%%%%%%%%%%
%%%%% EFFECTIVE Theory %%%%%%%%
%%%%%%%%%%%%%%%%%%%%%%%%
\begin{figure}[t]
%	\hspace*{-0.15in}
\begin{center}
	\includegraphics[width=3.4in]{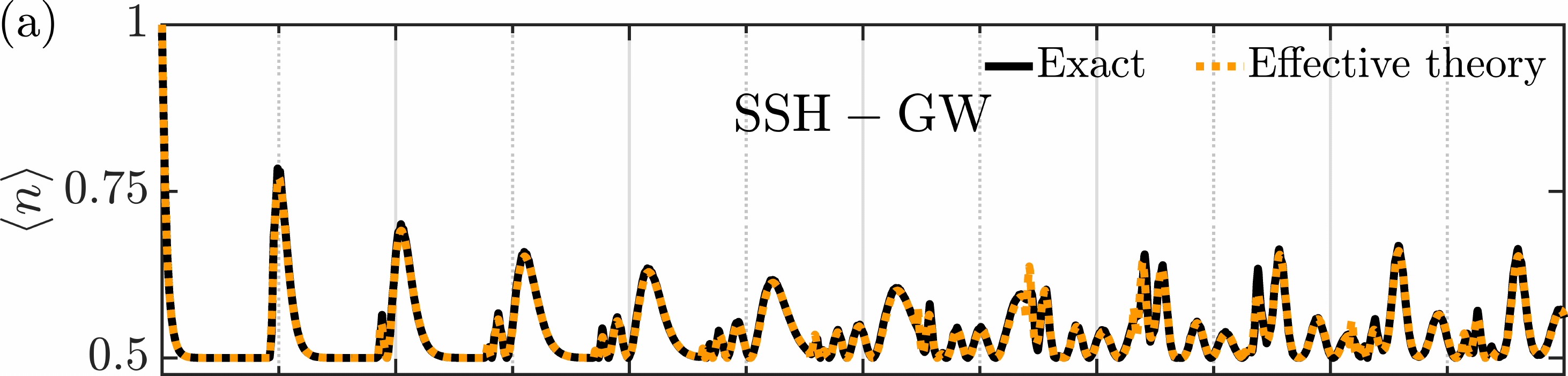}\\
    \includegraphics[width=3.4in]{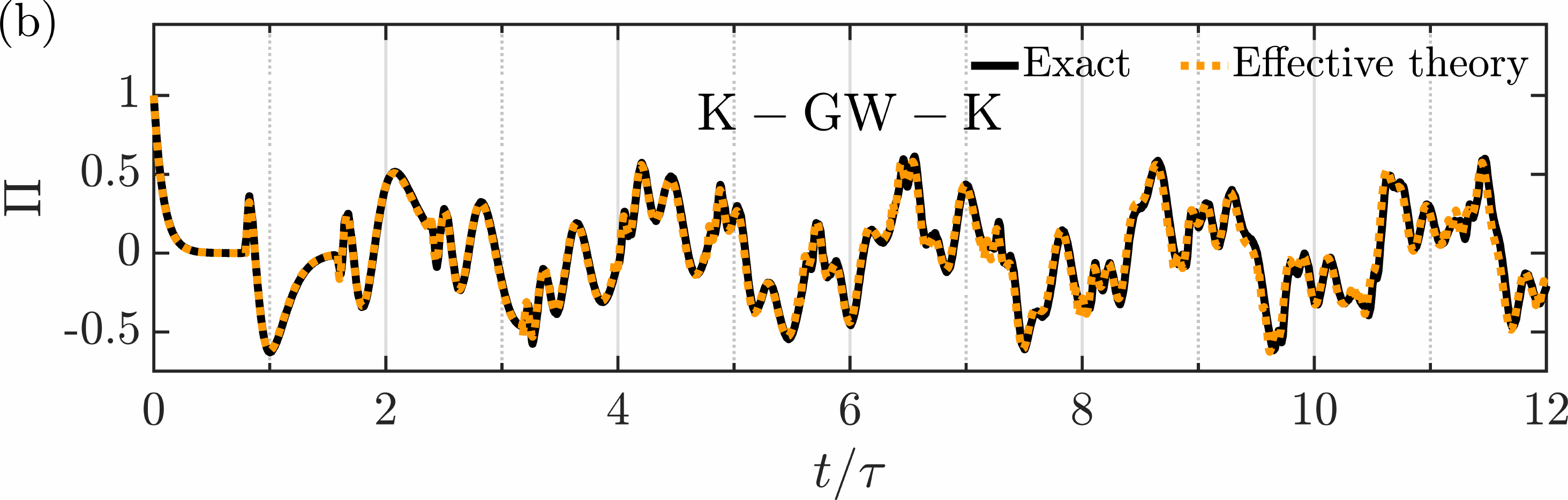}
	\vspace{-7mm}
	\caption{
		Comparison between the exact diagonalization (solid black lines) and the effective theory results (dotted orange lines). (a) Density at the rightmost site of the left SSH chain ($m_L=0.8w$, $N_L=50$), coupled via $\lambda_L=0.2w\,\theta(t)$ to a GW ($N=300$), with $\lambda_R=0$ for all $t$. (b) Fermion parity of the left Kitaev chain, for two Kitaev chains ($N_{L, R}=20$, $\Delta_{L, R}=-0.7w$) coupled via $\lambda_{L, R}=0.2w\,\theta(t)$ to a GW ($N=250)$. 
	} 
	\label{fig:exact-effective} 
	\end{center}
	\vspace{-8mm}
\end{figure}
%%%%%%%%%%%%%%%%%%%%%%%%

The exact solutions of Eq.~\eqref{eq:motion} are derived in Appendix~\ref{app:effective-theory}\,. For a single TBS $\hat\alpha_L$ coupled to $\hat\nu(x_L=0)$ via $\lambda_L=\lambda$, we find $\hat \alpha_L(t)=g_L(t)\hat \alpha_L(0) +\mathcal{X}_L[\hat \nu_0]$, where $\mathcal{X}_L$ is some functional of  $\hat \nu_0$ only, and
%%%%%
\begin{eqnarray}\label{eq:f}
g_L(t)=e^{-\Gamma t}\sum_{k=0}^{n}\tfrac{(-2\Gamma)^k }{k!}\sum_{j=k}^{n}
%x\tmultiset{n}{j-n} 
\tbinom{j-1}{k-1}
\zeta^j e^{\Gamma j\tau_r} (t- j \tau_r)^k,\quad\;
\end{eqnarray}
%%%%%
with $\Gamma=2|\lambda|^2/v_F$ and $n=\lfloor t/\tau_r \rfloor$.
%$n=\lfloor t/\tau_r \rfloor$.
 % and $\tmultiset{x}{y}$ the multiset coefficients. 
We can now identify the total revival time as $\tau \equiv \tau_r + \beta \tau_l$, where $\tau_l=1/\Gamma$ is the leakage time and $\beta \sim O(1)$ depends on the number of past reflections. In the low energy limit, $\Gamma\tau_r\gg 1$ and % one can see from Eq.~(\ref{eq:f}) that at 
at the $n$th revival cycle, the maximum amplitude scales like $|g_L^\text{max}| \sim n^{-1/3}$ for $n\gg1$ (see Appendix \ref{app:effective-theory} for a derivation). 

When two TBSs are coupled to $\hat\nu$ at $x_{L}=0$, $x_R=\ell$ with $\lambda_{L, R}=\lambda$, we obtain $\hat\alpha_a(t)=\sum_b g_{ab}(t)\hat\alpha_b(0)+\mathcal{Y}_{a}[\hat\nu_0]$, with $\mathcal{Y}_a$ is another functional of $\hat\nu_0$ only, and
%%%%%
\begin{eqnarray}
\label{eq:g-1}
&&g_{LL}(t) 
= g_{RR}(t)
= e^{-\Gamma t}\sum_{k=0}^{2n}\tfrac{(-2\Gamma)^k}{k!}\sum_{j=k}^{2n}
%\tmultiset{n}{j-n} 
\tbinom{j-1}{k-1}
\\
&&~\times
	\hspace{-0.4cm}
\sum_{{\hspace{0.1cm}m\geq j-n \geq 0}}
	\hspace{-0.2cm}
\tbinom{k}{2m} 
\left( \zeta e^{\Gamma \tau_r}\right)^{j-m}[t-(j-m)\tau_r]^k, \nonumber \\
\label{eq:g-2} 
&&g_{LR}(t)=\zeta g_{RL}(t)= e^{-\Gamma t}
	%\hspace{-0.2cm}
\sum_{k=0}^{2n'}
 	%\hspace{-0.2cm}
\tfrac{(-2\Gamma)^k}{k!}
	%\hspace{-0.2cm}
\sum_{j=k}^{2n'}
	%\hspace{-0.2cm}
%\tmultiset{n}{j-n} 
\tbinom{j-1}{k-1}
\\
&&~\times
	\hspace{-1cm}
\sum_{\hspace{0.7cm}m\geq j-n'-\frac{1}{2} \geq 0}
	\hspace{-0.9cm}
\tbinom{k}{2m+1}
e^{\frac{\Gamma \tau_r}{2}}
\left(\zeta e^{\Gamma \tau_r}\right) ^{j-(m+1)} 
[t-\tau_r(j-m-\tfrac{1}{2})]^k, \nonumber 
\end{eqnarray}
%%%%%
with $n'=\lfloor 
t/\tau_r+1/2\rfloor$.
% $n'=\lfloor t/\tau_r+1/2\rfloor$.
 These equations are matched against exact evolution results in Fig.~\ref{fig:exact-effective}\,, showing remarkable agreement.

\vspace{-2mm}
\myheading{Dynamically Induced Topology}%
Following the quench, the GW evolves into an intricate non-equilibrium state, and becomes entangled with the TBSs. So far, we have shown that, even though the GW is itself topologically trivial in equilibrium, its quenched state sustains signatures of coherent quantum statistics of TBSs over long times, a canonical signature of nontrivial topology. In the following we study other properties of the quench dynamics in the GW, which characterize the topological phase of the parent system supporting the TBSs. Specifically, these are: (i) propagation of quantum numbers in the GW; and (ii) patterns in the entanglement entropy. These properties are, in effect, temporal extensions of the topological properties of the systems hosting TBSs into the GW. Our analytical expressions for these calculations are presented in Appendix~\ref{app:trace-formula}.

\vspace{-2mm}
\mysubheading{Fermion Charge and Parity}%
Following the coupling of the SSH model to the GW, a disturbance in the density leaks into the GW. We study the charge carried by this disturbance by defining the smoothed charge operator~\cite{kivelson1982fractional} 
centered on site $r$, $\hat{Q}_r[h]=\sum_{r'} \mathsf{f}_{r-r'} \hat{n}_{r'}$, where $\mathsf{f}$ is a smoothing function, which we take to be  $\mathsf{f}_r= \exp(-r^2/l^2)$ with $l$ a smoothing length. Its expectation value and fluctuations are given by
%%%%%
\begin{eqnarray}
\hspace{-4mm}
Q_r(t) &=& \sum_{r'}  \mathsf{f}_{r-r'} [1-G_{r'r'}(t)] ,\\
\hspace{-4mm}
\delta Q_r^2(t) &=& \sum_{r'_1r'_2} \mathsf{f}_{r-r'_1} \mathsf{f}_{r-r'_2} G_{r'_1r'_2}(t)[\delta_{r'_1r'_2}-G_{r'_2r'_1}(t)].
\end{eqnarray}
%%%%%
Here $G_{rr'}(t) = \langle{\hat c\nodag_{r}\hat c_{r'}^\dagger}\rangle(t)$ is the equal time correlator between sites $r$ and $r'$, where $\langle{\hat\bullet}\rangle(t) = \langle{\Omega(t)}|{\hat\bullet}|{\Omega(t)}\rangle$ with $|{\Omega(t)}\rangle$ the state of the system at time $t$. 

%%%%%%%%%%%%%%%%%%%%%%%%
%%%%%%%%% FIG. 1 %%%%%%%%%%%
%%%%%% SSH CHARGE %%%%%%%%%%
%%%%%%%%%%%%%%%%%%%%%%%%
\begin{figure}
	%\centering \includegraphics[scale=0.97]{figs/fig1.pdf}
	\centering \includegraphics[width=3.4in]{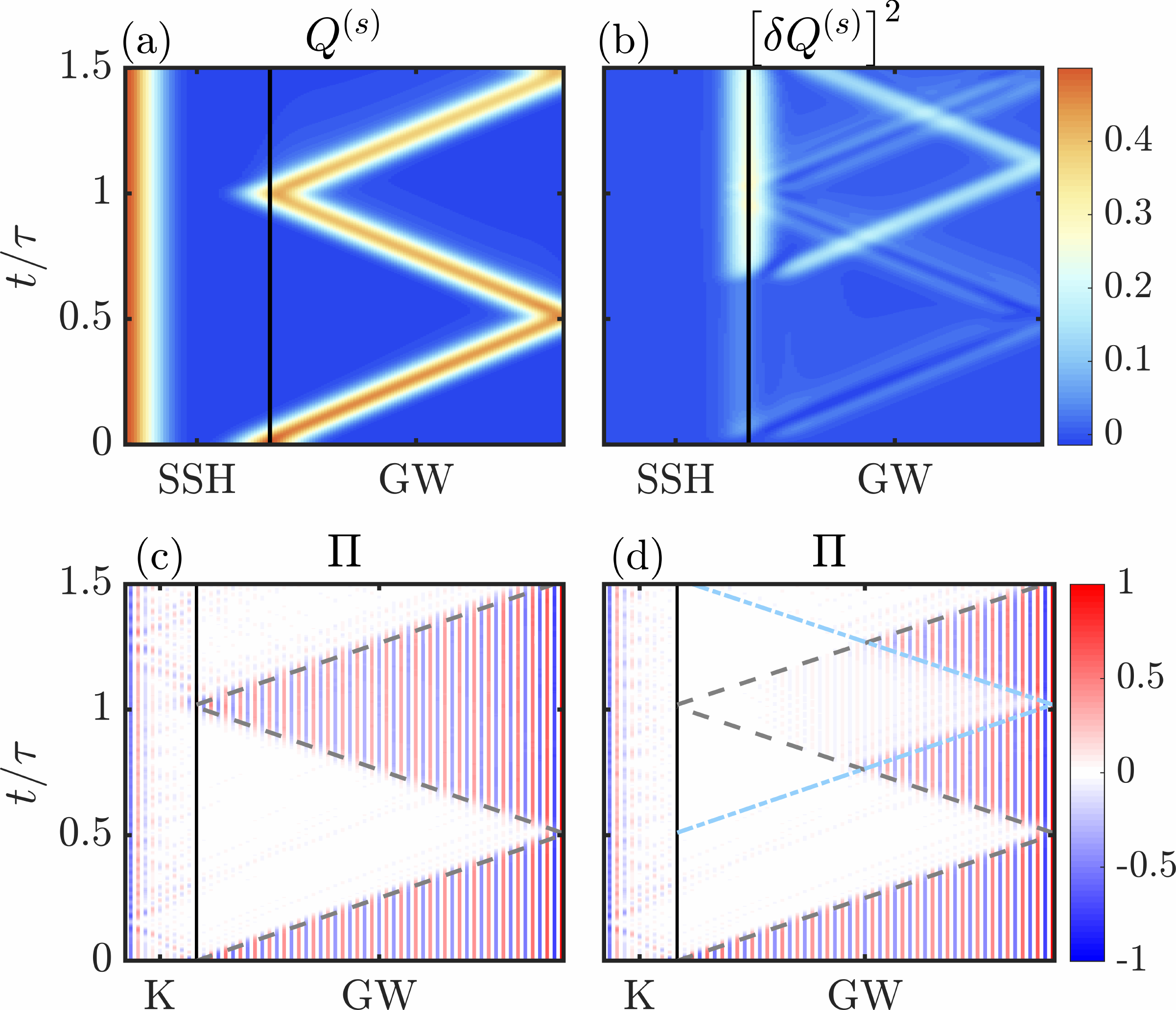}
	\caption{
		%Local charge and excess charge fluctuations for SSH-GW configuration, and parity propagation for K-GW configuration, following a quench.
		Local soliton charge (a) and excess fluctuations (b) for SSH-GW configuration following a double quench ($t_0=0$, $t_1\approx\tau/2$, $\lambda_L=0.5w$) of SSH ($N_L=50$, $m_L=0.8w$) and GW ($N=100$) with a smoothing length $l=20$. The propagating excitation carries a quantized charge $1/2$ and contributes negligibly to the charge fluctuations compared to the soliton-free fermi sea. Fermion parity to the left of a cut for K-GW configuration with a Kitaev chain ($N_L=20$, $\Delta_L=-0.3w$) and a GW ($N=100$) following a single quench ($t_0=0$, $\lambda_L=w$) (c) and a double quench ($t_0=0$, $t_1\approx\tau/2$, $\lambda_L=w$) (d). The dashed lines indicate propagation at $v_F$ following the quenches. 
	}
	\label{fig:charge} 
	\vspace{-5mm}
\end{figure}

To isolate the effects of the SSH soliton in the GW, we take the soliton charge and charge fluctuations as $Q^{(s)} \equiv Q^{(+)}-Q^{(-)}$ and $[\delta Q^{(s)}]^2 \equiv [\delta Q^{(+)}]^2 - [\delta Q^{(-)}]^2$, 
where the superscript $+$ ($-$) indicates that the initial state of the system has (no) solitons. Calculated in this manner, $Q^{(s)}$  and $[\delta Q^{(s)}]^2$ are plotted in Fig.~\ref{fig:charge}$\,$ as functions of the position $r$ and time $t$. For simplicity, we have taken $\delta Q^{(-)}(t)=\delta Q^{(+)}(t=0)$, since without a soliton the time-dependence of the charge fluctuations leaking into the GW is negligible. As depicted in Fig.~\ref{fig:charge}, as the soliton propagates in the GW, the fractional charge $Q^{(s)}=1/2$ \cite{jackiw1976solitons} moves with it. Furthermore, at the center of the moving soliton the charge fluctuations $\delta Q^{(s)}\approx 0$, signifying that the moving fractional charge is, to a good approximation, a good quantum number. 

For quenches involving the Kitaev chain, the density contains no signatures of the Majorana mode. Instead, we look at the fermion parity operator. We define the anomalous correlator $F_{rr'}(t)=\langle \hat c_r \hat c_{r'} \rangle$ and combine $G$ and $F$ to form the Nambu correlator 
%%%%% PRL $
\begin{equation}
\mathcal{G} = \begin{pmatrix} G & F \\ F^\dagger & \id - G^\trans \end{pmatrix}.
\end{equation}
%%%%% PRL $
In Appendix~\ref{app:trace-formula} we show that, without interactions, the parity $\Pi_{\mathsf{s}}$ for a subsystem with $N_\mathsf{s}$ sites is given by a simple and useful formula,
%%%%%
\begin{align}
	\Pi_{\mathsf{s}} = (-1)^{\frac12{N_\mathsf{s}(N_\mathsf{s}-1)}}\pf\left[ \Sigma_x(\id-2\mathcal{G}_\mathsf{s}) \right],
\end{align}
where $\mathcal{G}_\mathsf{s}$ is obtained by a straightforward projection to the sites of $\mathsf{s}$. The results following a single and a double quench, are displayed in the lower panels of Fig.~\ref{fig:charge}$\,$, where $\mathsf{s}$ comprises all the sites to the left of a cut.

We find that following the quench the Majorana TBS propagates through the GW, rendering the parity of any subsystem of the segment between the TBS and its remote partner zero. The same behavior occurs in a double quench when tunneling is turned off at $t_1$, indicative of a Majorana pair creation at the interface: one is trapped by the Kitaev chain while the other starts to propagate through the GW. The subsystem parity through a cut remains zero for any site between these new partners. 

Interestingly, the double quench for the SSH solitons shows signatures of local charge fluctuations after $t_1$ not of charge itself. This can also be understood as a pair creation but now of particle-hole excitations, one trapped to the SSH chain and the other propagating through the GW. We now examine this scenario through the entanglement of subsystems.

\vspace{-3mm}
\mysubheading{Entanglement Entropy}%
Following the quench the state of the TBSs (thus, the system hosting it) and the GW become entangled. Thus, we expect that the propagation of TBSs and the concomitant fluctuations must be accompanied with the propagation of entanglement in the GW. 

The von Neumann entanglement entropy of the subsystem $\mathsf{s}$ with reduced density matrix $\hat\rho_{\mathsf{s}}$ is $S_\mathsf{s} = \tr(\hat\rho_{\mathsf{s}}\log\hat\rho_{\mathsf{s}})$. In the non-interacting GW, one can use trace formulae \cite{klich2014note} to obtain~\cite{sen2016entanglement} (see Appendix~\ref{app:trace-formula} for details):
%%%%%
\begin{equation}\label{eq:ent-G}
S_{\mathsf{s}} = -\frac{1}{2} \tr \left[\mathcal{G}_{\mathsf{s}}\log \mathcal{G}_{\mathsf{s}}+(\id-\mathcal{G}_{\mathsf{s}})\log(\id-\mathcal{G}_{\mathsf{s}})\right].
\end{equation}
%%%%%
It was suggested in \cite{klich2008quantum} that the entanglement entropy can be related to the counting statistics of the number operator through the generating function
$ \chi_\mathsf{s}(\phi)$. For a normal system, i.e. when $F=0$, $S_\mathsf{s}=\sum_{m}\frac{\alpha_{m}}{m!}Q^{(m)}_{\mathsf{s}}$, where $Q_\mathsf{s}^{(m)} \equiv {d^m \log\chi_\mathsf{s}}/{d(i\phi)^m}\vert_{\phi=0}$ are the cumulant moments of the number operator and $\alpha_{m}$ are given by an integral form; the first few terms are
%%%%% PRL $
\begin{equation}
S_\mathsf{s}=\frac{\pi^2}{3}Q^{(2)}_{\mathsf{s}}+\frac{\pi^4}{45}Q^{(4)}_{\mathsf{s}}+\frac{2\pi^6}{945}Q^{(6)}_{\mathsf{s}}+\cdots.
\end{equation}
%%%%% PRL $
In a paired state, the same relationship holds if one takes $\chi$ to generate the counting statistics of Bogoliubov quasiparticles instead of the original paired particles. This can be understood by noting that in the quasiparticles basis, $\mathcal{G}$ is diagonal, and thus $F=0$ as before. We note, however, that once projected to a subsystem this basis is not the same as the quasiparticle basis of the entire system. In other words, the two operations of diagonalizing $\mathcal{G}$ and projecting to the subsystem do not commute. We do not explore the subtleties of this relationship here. Instead, we report our numerical results for the quench dynamics entanglement entropy and particle (as opposed to quasiparticle) fluctuations. We find that $S_\mathsf{s} \sim \frac{\pi^2}3 Q^{(2)}_{\mathsf{s}}$ \cite{hsu2009quantum,song2010general,song2011entanglement,song2012bipartite,herviou2017bipartite} holds in all cases.

%%%%%%%%% FIG. 3 %%%%%%%%%%%
%%%%%% ENTANGLEMENT%%%%%%%%
%%%%%%%%%%%%%%%%%%%%%%%%
\begin{figure}[t]
\begin{center}
	\includegraphics[width=3.4in]{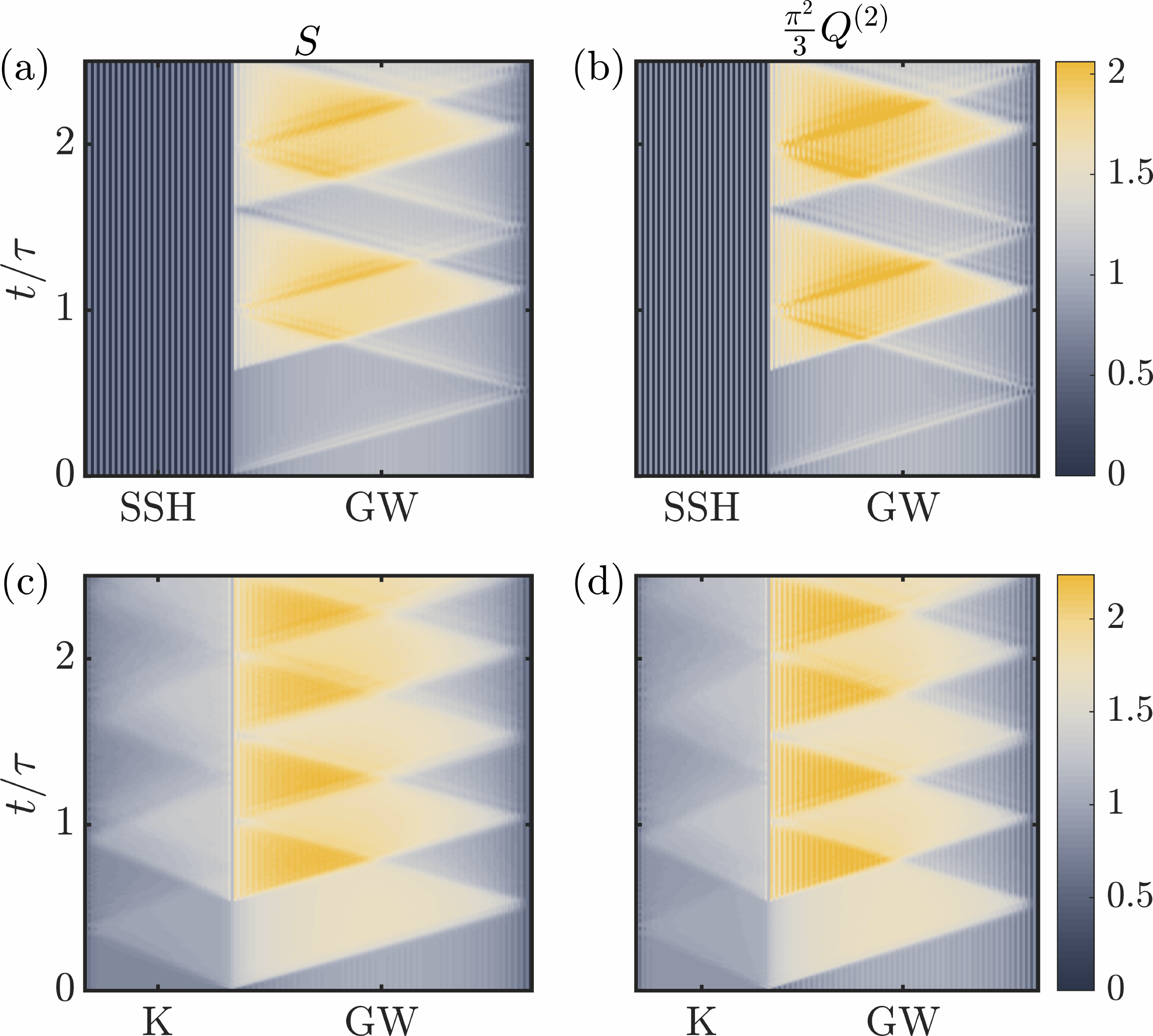}
	\caption{
		Entanglement entropy (left) and charge fluctuations (right) through a cut following a double quench, for SSH-GW (upper, $N_L=50$, $m_L=0.8w$, $\lambda_L=0.5w$) and Kitaev-GW (lower, $N_L=50$, $\Delta_L=-0.3w$, $\lambda_L=w$) configurations, as function of time, with $t_0=0$ and $t_1\approx\tau/2$, for %coupling $\lambda_L=0.5w$ and
		$N=100$.
	}
	\label{fig:entanglement} 
\end{center}
	\vspace{-5mm}
\end{figure}

When comparing the results of the SSH and the Kitaev cases, several distinct behaviors of the entanglement entropy occur, see Fig.~\ref{fig:entanglement}$\,$. It is either finite over a small window around the $v_F$ cone (SSH case, following $t_0$) or displays a pattern which extends over the entire spatial path (Kitaev case, following $t_0$). Following $t_1$ in a double quench, it displays the extended pattern for both cases. 

For the SSH solitons, the reason for the change in behavior for $t>t_1$ can be related to the presence of an unmatched TBS with fractional charge $1/2$ within the disconnected GW. To allow for its presence, the number fluctuations on the \emph{entire} GW must increase. This is consistent with the local charge and its fluctuations, Fig.~\ref{fig:charge}\,a and Fig.~\ref{fig:charge}\,b following $t_1\simeq \tau/2$, which we explained by a particle-hole pair creation.

For the Majorana TBS, following a single quench, the charge fluctuations account for the maximal uncertainty in parity through any cut between the TBS and its remote partner. Following a double quench, the behavior is again consistent with the generation of a pair of two Majorana TBSs near the interface, one trapped and one propagating.

\myheading{Discussion}%
Proximity quenches of topological and gapless wires offer a possibility to inject topological bound states into the wire, which behave as mobile anyons augmenting the properties of the wire over a large coherence time. The dissipation in the GW occurs through loss of amplitude of the anyon as it disperses in the gapless medium, while its phase continues to bear the imprint of the parent TBS. The power-law decay of the maximum amplitude $\sim n^{-1/3}$ after $n\gg1$ revivals may be expected since there are no timescales in the GW left in this limit. The slow decay demonstrates the long-lived coherence time of the dynamically induced topology. 

\begin{figure}
	\includegraphics[width=0.49\textwidth]{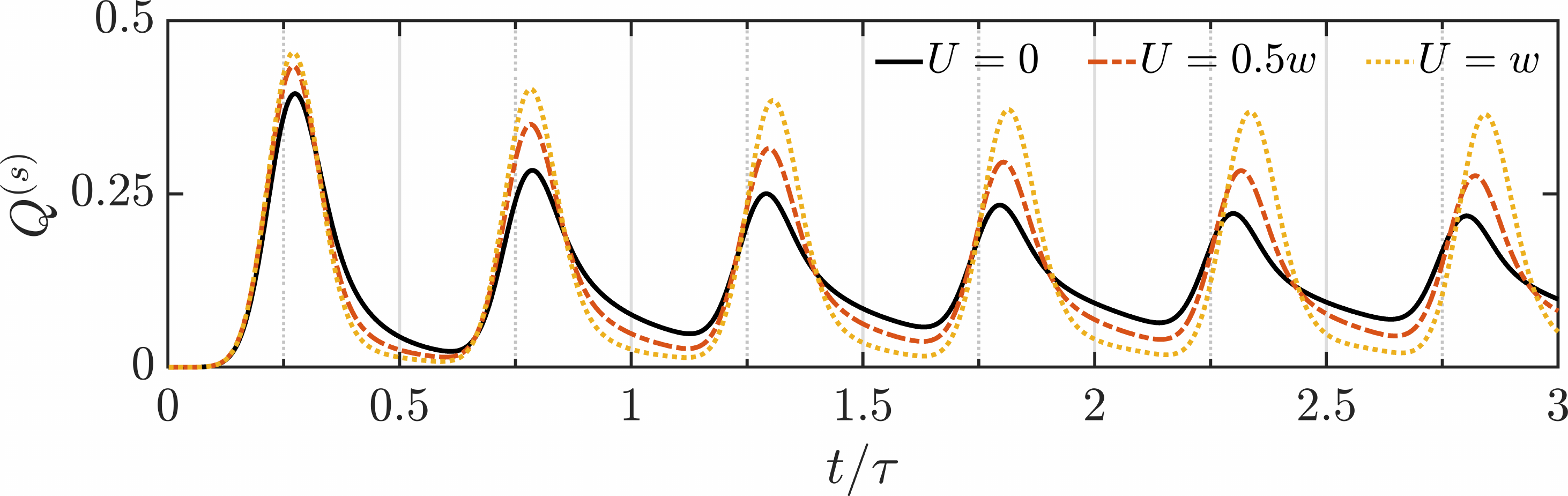}
	\caption{Soltion charge $Q^{(s)}$ in the presence of interactions. The charged is measured within a window centered at the middle of the GW.  Here $N=140$, $N_L=20$, $m_L =0.8w$, $\lambda_L=w$ and smoothing length $l=20$.}
	\label{fig:InteractionsSSH} 
\end{figure}

We find further signatures of the dynamically induced topology in the GW in the charge, fermion parity, charge fluctuations, and the entanglement entropy. In the presence of interactions~\cite{sela2011majorana,fidkowski2012universal,affleck2013topological}, the GW is effectively described by the Luttinger liquid theory with a renormalized Fermi velocity
%%%% -- PRL $
\begin{equation}
v = v_F \frac{\pi \sqrt{1-\tilde{u}^2}}{2\arccos(\tilde{u})},
\end{equation}
%%%% -- PRL $
where $\tilde{u}\equiv u/v_F$~\footnote{See, e.g., M.~Takahashi, Thermodynamics of one-dimensional solvable problems,  Cambridge University Press, (1999).}, see Appendix~\ref{app:robustness} for details. The conformal field theory at discrete values of the Luttinger parameter, $1/K= 2-\frac2\pi\arccos\tilde{u}\in\mathbb{N}$, can be embedded within an orbifold theory~\cite{hsu2009quantum}, but appropriate BCOs can be found for other values of $K$ \cite{affleck1994fermi}.
Here, we explore the effect of interactions numerically using tDMRG. For the SSH case, we study the soliton charge within a static window around the center of the lead as function of time, see Fig.~[\ref{fig:InteractionsSSH}].
To counteract the effect of renormalized Fermi velocity, we measure the time in units of a renormalized return time $\tau_r v_F/v$. Remarkably, we find that in addition, interactions can significantly decrease the decay of the soliton charge and lead to sharper peaks. We discuss the GW in the presence of disorder in Appendix~\ref{app:robustness}.

%%%%%%%%% FIG. 9 %%%%%%%%%%%
%%%%% SSH INTERACTIONS %%%%%%%
%%%%%%%%%%%%%%%%%%%%%%%%
%%%%%%%%%%%%%%%%%%%%%%%%

Quantities such as the (R\'enyi) entanglement entropy and, by a straightforward extension, the fidelity, are measurable using techniques developed in Ref.~\cite{islam2015measuring}. The extension required here is to quench two copies of the system at different times, and let them interfere. Such an extension could pave the way to measurement of the dynamically induced topology, including measurement of signatures of non-abelian statistics through fidelity revivals and propagation of entanglement with a unique behavior. In addition, if the two systems are not identical but instead admit different initial occupations of the fermion TBSs (or different parities of the superconductors), the interference fringes would be shifted by the relative statistical phase, Eq.~\ref{eq:stat-phase}$\,$, unveiling the fractional statistics. Our work opens the door to study the localization and thermalization properties of anyons, with and without interactions. It is also of interest to consider extensions to ``anyon trains'' and to interacting TBSs, e.g. parafermions \cite{fradkin1980disorder,fendley2014free} in future.

\vspace{-5mm}
\begin{acknowledgements}
\vspace{-3mm}
\noindent DD and EG acknowledge useful discussions with D.~Ariad, D.~Cohen, M.~Schechter and E.~Sela. BS acknowledges useful conversations with M. Kolodrubetz. DD and EG acknowledge support from the Israel Science Foundation under grant
No. 1626/16. EG and BS acknowledge support from the Binational
Science Foundation through grant No. 2014345. This work is supported in part by the NSF CAREER grant DMR-1350663 as well as the College of Arts and Sciences at Indiana University. This work was performed in parts at the Aspen Center for Physics, supported by the NSF grant PHY-1607611.
\end{acknowledgements}

%\setcounter{figure}{0}
%\setcounter{equation}{0}
%\makeatletter
%\renewcommand{\thefigure}{S\@arabic\c@figure}
%\renewcommand{\theequation}{S\@arabic\c@equation}
%\makeatother

%%%%%%%%%%%%%%%%%%%%%%
%%%  Exact Diagonlization %%%
%%%%%%%%%%%%%%%%%%%%%%

%\newpage 

\appendix
%\section{Supplemental Material}
%
%In the Supplemental Material we provide additional information regarding our numerical calculations and provide proofs of the key formulas from the main text. We introduce the lattice models behind the numerical evolution we perform. We derive the formulas for the entanglement entropy, the full counting statistics and the fermion parity. We describe the Onishi formula which is used to calculate the overlap of states and extract the fidelity and phase. We describe the low energy effective theory, derive the dynamics of topological bound states, and present the correlators of boundary changing operators. Finally, we discuss the robustness of our results in the presence of interactions and disorder.

\section{Exact time evolution}\label{app:exact-time-evolution}

In the main text we considered two canonical lattice models, namely the SSH and the Kitaev models, that support, respectively, regular and Majorana fermion TBSs in their topological phases. As is well known, the SSH TBSs are solitons that carry a fractional charge $1/2$, whereas the Majorana TBSs in the Kitaev model are neutral. However, Majorana TBSs implement the non-Abelian fractional statistics of Ising anyons. In this Appendix, we provide the details of our numerical method for the exact diagonalization of these models for the non-interacting GW (with or without disorder).

Our time-dependent model Hamiltonian in the main text is $\hat H(t)=\hat H_{L}+\hat H_{\text{GW}}+H_{R}+\hat H_{\text{t}}(t)$. 
The topological wire Hamiltonian, for $a\in\{L,R\}$, is
%%%%%
\begin{align}\label{eq:sys}
\hat H_{a} = \sum_{s=1}^{N_{a}-1}\left[ w\nodag_{a,s} \hat d_{a,s}^{\dagger} \hat d\nodag_{a,s+1}+ \Delta \hat d_{a,s} \hat d_{a,s+1}+\text{h.c.}\right], 
\end{align}
%%%%%
where $\hat d_{a,s}^\dag$ creates a fermion in the $a$ wire at system site $s$, $w_{a,s} = w_a+(-1)^s m_a$ is the dimerized hopping amplitude with $m_a$ the dimerization strength, and $\Delta_a$ is the (p-wave) pairing amplitude. For $m_a\neq0, \Delta_a=0$ we obtain the SSH model; for $m_a=0,\Delta_a\neq 0$, we obtain the Kitaev model. 
The GW Hamiltonian is 
%%%%%
\begin{align}\label{eq:GW}
\nonumber \hat H_{\text{GW}} &= \hat{H}_f+\hat{H}_{\text{dis}}+\hat{H}_\text{int}\\ 
&=~ w\sum_{r=1}^{N-1}\left(\hat c_{r}^{\dag}\hat c\nodag_{r+1} + \text{h.c.} \right)
+ \sum_{r=1}^N \mu_r\left(\hat n_r -\frac{1}{2} \right) \nonumber\\
&\quad + u\sum_{r=1}^{N-1}\left( \hat n_r-\frac{1}{2}\right)\left(\hat n_{r+1}-\frac{1}{2}\right),
\end{align}
%%%%%
where $N$ is the number of sites, $\hat c_r^\dag$ creates a fermion at lead site $r$, $\hat n_r = \hat c_r^\dag \hat c\nodag_r$ is the number operator, $w$ is the hopping amplitude, $u$ is the short-range interaction strength and $\mu_r$ is the potential disorder with strength $W$. In our numerics, we draw $\mu_r$ from a uniform distribution over  $[-W/2,W/2]$. 
Lastly, the tunneling Hamiltonian
%%%%%
\begin{equation}
\hat H_{\text{t}}(t) = \sum_{r,a,s} \lambda_{r;a,s}(t)  \hat d_{a,s}^{\dagger} \hat c\nodag_{r} + \text{h.c.},
\end{equation}
%%%%%
where $\lambda_{r;a,s}(t)$ is the tunneling amplitude between system $a$ site $s$ and GW site $r$. The initial state of the system is taken to be the ground state $|\Omega\rangle$ of the uncoupled system and GW at half filling.

Working in the Nambu basis, let $\hat C^\trans=(\hat c^\trans, \hat c^\dagger)$ where $\hat c^\trans=(\hat c_{1}, \ldots \hat c_{N})$ in some (say, the position) basis, and $\hat \Gamma^\trans=(\hat \gamma^\trans,\gamma^{\dagger})$ with $\hat \gamma^\trans=(\hat \gamma_{1},\ldots,\hat \gamma_{N})$ the quasi-particle operators related to the original basis by the unitary Bogoliubov transformation,
%%%%%
\begin{equation}
\hat C=\mathcal{W} \hat \Gamma,\quad
\mathcal{W}=\begin{pmatrix}
U & V^{*}\\
V & U^{*}
\end{pmatrix}.
\end{equation}
%%%%%
The matrix $\mathcal{W}$ diagonalizes the BdG Hamiltonian
$\mathcal{H}_{\text{d}}=\mathcal{W}^{\dagger}\mathcal{H}\mathcal{W}=\text{diag}(\epsilon_{1},\ldots,\epsilon_{N},-\epsilon_{1},\ldots,-\epsilon_{N})$
at $t=0$, with the positive eigenvalues $\epsilon_\alpha>0$ associated
with the first $N$ columns of $\mathcal{W}$. 

The ground state of the entire system $|\Omega\rangle$ is defined as the unique state annihilated by all quasiparticle operators,
%%%%%
\begin{equation}
\gamma_{\alpha}|\Omega\rangle=0,\quad\text{for all }\epsilon_\alpha>0.
\end{equation}
%%%%%
One can construct this state explicitly as $\prod_{\epsilon_\alpha>0}\gamma_{\alpha}|0\rangle$. When $\inner{0}{\Omega}\neq0$, where $\ket{0}$ is the vacuum state annihilated by $\hat c$, the ground state has the Thouless representation,
%%%%%
\begin{equation}
\ket{\Omega}=A\exp\left[\frac{1}{2} \hat c^\dagger Z \hat c^{\dagger\trans} \right]\ket{0},
\end{equation}
%%%%%
where $A$ is a normalization factor and $Z = - (VU^{-1})^\dagger$. %$Z = -G^{-1}F$.

The Nambu Green's function is defined as $\mathcal{G} = \bra{\Omega} \hat C \hat C^\dagger \ket{\Omega}$. Using the above relations and rotating to the quasiparticle basis, where $\bra{\Omega} \hat\Gamma \hat\Gamma^\dagger \ket{\Omega} = (\id + \Sigma_z)/2$, we find $\mathcal{G} = \begin{pmatrix} G & F \\ F^\dagger & \id - G^\trans \end{pmatrix}$, with
%%%%%
\begin{align}
G &= \bra{\Omega} \hat c \hat c^\dagger \ket{\Omega} = UU^\dagger, \\
F &= \bra{\Omega} \hat c \hat c^\trans \ket{\Omega} = UV^\dagger.
\end{align}
%%%%%
Note that the kernel of the exponent in the Thouless representation of the ground state can be written as $Z=-G^{-1}F$. The unitarity of $\mathcal{W}$ yield the following relations: $G(\id-G) = FF^\dagger$ and $GF = FG^\trans$. We note that these are true for the entire system and do not apply to projected forms of $G$ and $F$. We can also see easily that the matrix $2\mathcal{G}-\id$ (and its projections to a subsystem) are particle-hole symmetric:
%%%%%
\begin{equation}
\Sigma_x (2\mathcal{G} - \id)^\trans \Sigma_x = -(2\mathcal{G} - \id).
\end{equation}
%%%%%
Equivalently, the matrix $\Sigma_x(2\mathcal{G}-\id)$ and its subsystem projections are anti-symmetric.

The evolution for times $t>0$
is performed using the unitary evolution operator $\mathcal{U}(t)= \Texp\left[-i\int_{0}^{t}\mathcal{H}(s)ds\right]$,
%%%%%
\begin{equation}
\hat C(t)=\mathcal{U}(t) \hat C(0)=\mathcal{U}(t)\mathcal{W} \hat \Gamma(0).
\end{equation}
%%%%%
We may also find $\ket{\Omega(t)}$ in the Thouless form with $Z(t) = \mathcal{U}(t) Z(0) \mathcal{U}^\dagger(t)$ with the equal time Green's function $\mathcal{G}(t)=\mathcal{U}(t)\mathcal{G}(0)\mathcal{U}^{\dagger}(t)$.

\section{Overlaps, Fidelity and Phase}\label{app:overlap-fidelity-phase}

The overlap between two states in the Thouless representation, 
\begin{equation}
%|\Omega_{\alpha}\rangle=A_{\alpha}\exp\left[\frac{1}{2}\sum_{ij}Z_{ij}^{(\alpha)}\hat c_{i}^{\dagger}\hat c_{j}^{\dagger}\right]|0\rangle,\quad\alpha=1,2,
|\Omega_{\alpha}\rangle=A_{\alpha}\exp\left[\frac{1}{2}\hat c^{\dagger} Z^{(\alpha)} \hat c^{\dagger\trans}\right]|0\rangle,\quad\alpha=1,2,
\end{equation}
is
\begin{equation} \label{eq:overlap}
\langle\Omega_{1}|\Omega_{2}\rangle=s_{N}A_{1}A_{2}\pf(\mathcal{Q}),
\end{equation}
where $s_{N}=(-1)^{N(N+1)/2}$ and
\begin{equation}
\mathcal{Q}=\left(\begin{array}{cc}
Z^{(2)} & - \id\\
 \id & -Z^{(1)*}
\end{array}\right).
\end{equation}
To prove this we will apply the formalism of fermion coherent states,
defined as $|\xi\rangle=e^{-\hat c^{\dagger} \xi}|0\rangle=\prod_{k=1}^{N}(1-\xi_{k} \hat c_{k}^{\dagger})|0\rangle$,
where $\xi = (\xi_{1},\ldots,\xi_{N})^\trans$ and its conjugate $\xi^*$ are $2N$ anti-commuting independent Grassmann numbers. In this language, we write the resolution of identity 
$
\id=\int d\mu(\xi) |\xi\rangle\langle\xi|,
$
the measure 
$
d\mu(\xi)=\prod_{i}\left(d\xi_{i}^{*}d\xi_{i}\right)e^{-\sum_{i}\xi_{i}^{*}\xi_{i}},
$
and the trace of an operator,
$
\tr\hat{A} =\int d\mu(\xi)\langle-\xi|\hat{A}|\xi\rangle.
$
We also note that when we consider a fermion coherent state $|\xi\rangle$ and apply an operator of the form $\hat{O}_{M}=e^{-\hat c^{\dagger}M\hat c}$, then $\hat{O}_{M}|\xi\rangle=|e^{-M}\xi\rangle$ is satisfied. Using the above, we can calculate the overlap

\begin{align}
\langle\Omega_{1}|\Omega_{2}\rangle&=A_{1}A_{2}\int d\mu(\xi)\exp\left(\frac{1}{2} \xi^\trans Z^{(1)*} \xi\right)\nonumber \\
&\quad\times\exp\left(\frac{1}{2}\xi^{*\trans}Z^{(2)}\xi^{*}\right),
\end{align}
which can be rewritten in matrix form as
\begin{align}
\langle\Omega_{1}|\Omega_{2}\rangle & =A_{1}A_{2}\int\prod_{i}d\xi_{i}^{*}d\xi_{i}\exp\left(\frac{1}{2}\Xi^\trans\mathcal{Q}\;\Xi\right),
\end{align}
where $\mathcal{Q}$ is as defined above and the Grassmann spinor $\Xi^\trans=(\xi^{*\trans},\xi^\trans)$.
The anti-symmetric matrix $\mathcal{Q}$ can always be transformed
into canonical form by means of an orthogonal transformation $\mathcal{Q}=M\mathcal{Q}_{c}M^\trans,$
\begin{align}
\mathcal{Q}_{c} & =\left(\begin{array}{cc}
0 & \Lambda\\
-\Lambda & 0
\end{array}\right),
\end{align}
where $\Lambda=\mbox{diag}(\lambda_{1},\ldots,\lambda_{N})$ with
$\lambda_{1},\ldots,\lambda_{N}$ non-negative real numbers. In terms
of the spinor $\Psi=M\Xi$ 
\begin{align}
& \langle\Omega_{1}|\Omega_{2}\rangle\nonumber \\
& =A_{1}A_{2}\det M\int\prod_{i}d\psi_{i}^{*}d\psi_{i}\exp\left(\sum_{i=1}^{N}\lambda_{i}\psi_{i}^{*}\psi_{i}\right)\nonumber \\
& =(-1)^{N}A_{1}^{*}A_{2}\det M\det\Lambda.
\end{align}
Using the relation
\begin{equation}
\det(\Lambda)=(-1)^{N(N-1)/2}\pf\left(\begin{array}{cc}
0 & \Lambda\\
-\Lambda^\trans & 0
\end{array}\right),
\end{equation}
we can write, using $\pf(\mathcal{M}\mathcal{A}\mathcal{M}^\trans)=\det(\mathcal{M})\pf(\mathcal{A})$,
\begin{align}
\langle\Omega_{1}|\Omega_{2}\rangle & =A_{1}A_{2}\det (M)\pf(\mathcal{Q}_{c})\nonumber \\
& =s_{N}A_{1}A_{2} \pf(\mathcal{Q}),
\end{align}
which coincides with Eq.~\eqref{eq:overlap}.

The Onishi formula states that the fidelity $\mathcal{F}$ is given by
\begin{equation} \label{eq:onishi}
\mathcal{F}=|\langle\Omega_{1}|\Omega_{2}\rangle|^{2}=\left|\det\left(U_{1}^{\dagger}U_{2}+V_{1}^{\dagger}V_{2}\right)\right|.
\end{equation}
To prove it, we use Eq.~\eqref{eq:overlap} and the relation $\pf^2 A=\det A$, so up
to a sign $s$,
\begin{equation}
\langle\Omega_{1}|\Omega_{2}\rangle=sA_{1}A_{2}\sqrt{\det\left(\begin{array}{cc}
	Z^{(2)} & - \id\\
	 \id & -Z^{(1)*}
	\end{array}\right)}.
\end{equation}
Using the relation $\det\left(\begin{array}{cc}
A & B\\
C & D
\end{array}\right)=\det(AD-BC)$ when $C$ and $D$ commute and noting the Sylvester's determinant
theorem $\det( \id+AD)=\det( \id+DA)$
\begin{align}\nonumber
\langle\Omega_{1}|\Omega_{2}\rangle=s(-1)^{N(N+1)/2}A_{1}A_{2}\sqrt{\det\left( \id-Z^{(2)}Z^{(1)*}\right)}.
\end{align}
By substituting our relations for $Z^{(\alpha)}$ and using their anti-symmetric property, we can rewrite this expression
\begin{align}
& \langle\Omega_{1}|\Omega_{2}\rangle =sA_{1}A_{2}\frac{\sqrt{\det\left(U_{2}^\trans U_{1}^{*}+V_{2}^\trans V_{1}^{*}\right)}}{\sqrt{\det U_{2}\det U_{1}^{*}}}.
\end{align}
Taking $|\Omega_{1}\rangle=|\Omega_{2}\rangle$, we express the normalization
for $\alpha=1,2$ as
$1 =\langle\Omega_{\alpha}|\Omega_{\alpha}\rangle=sA_{\alpha}^{2}/|\det U_{\alpha}|$, so we choose $s=1$ and get $A_{\alpha}=\sqrt{|\det U_{\alpha}|}$. The overlap
is therefore
\begin{align}
\langle\Omega_{1}|\Omega_{2}\rangle=e^{-\frac{i}{2}\arg\det U_{2}-\frac{i}{2}\arg\det U_{1}^{*}}\sqrt{\det\left(U_{1}^{\dagger}U_{2}+V_{1}^{\dagger}V_{2}\right)},
\end{align}
which coincides with Eq.~\eqref{eq:onishi} after multiplying each side by its complex conjugate.

\section{Low energy effective field theory}\label{app:effective-theory}
In this Appendix, we derive the low energy effective theory and the dynamics of topological bound states. We them present expressions for the correlators of BCOs. 

In the low-energy theory the GW is described by an effective theory obtained by linearizing the lattice model $H_f$ in Eq.~\eqref{eq:GW} near the Fermi wave-vector $k_F$ and taking the continuum limit where the lattice constant $\mathsf{a}\ll \ell=(N+1)\mathsf{a}$. In the following we set $\mathsf{a}=1$.
We start from momentum space representation
$\hat{H}_{f}= -2w\sum_{\tilde{k}} \cos(\tilde{k}) 
\hat n_{\tilde{k}}
%c^\dagger_{\tilde{k}}c_{\tilde{k}}
$, 
and linearize near the Fermi points: for right ($+$) movers $\tilde{k} =k_F+k=\frac{\pi}{2}+k$ and for left ($-$) movers 
$\tilde{k}=-k_F+k=-\frac{\pi}{2}+k$. For small $k$ we obtain:
\begin{align}\label{H_free}
\nonumber \hat H_f&= -2w
\hspace{-0.4cm}
\sum_{-\Lambda<k<\Lambda}
\hspace{-0.4cm}
\left[  
\cos \left(\frac{\pi}{2}+k\right)
\hat n_{+,k}
%c^\dagger_{+k}c_{+k}
+\cos\left(-\frac{\pi}{2}+k\right)
\hat n_{-,k}
%c^\dagger_{-k}c_{-k}
\right]\\ 
&\approx 2w
\hspace{-0.4cm}
\sum_{-\Lambda<k<\Lambda}
\hspace{-0.4cm}
k\left[  
\hat n_{+,k}
-
\hat n_{-,k}
%c^\dagger_{+k}c_{+k}
%-c^\dagger_{-k}c_{-k}
\right] .
\end{align}
where $\Lambda\sim 1/\mathsf{a}$ is a high momentum cutoff, and $\hat n_{\pm,k}=\hat c^\dag_{\pm,k} \hat  c_{\pm,k}$ are the momentum mode occupations near $\pm k_F$. In the same limit we define the real space fields
\begin{align}
\hat \psi_\pm(x)&=\frac{1}{\sqrt{\ell}}\sum_{k}e^{\pm ikx}\hat c_{\pm, k},
%\psi_-(x)&=\frac{1}{\sqrt{L}}\sum_{k}e^{-ikx}c_{-k},
\end{align}
where the Fermi velocity is $v_F=2w\sin (k_F) = 2w$ and  the Hamiltonian describing the fields is
\begin{align}\label{Lut1}
\hat{  H}_f&=v_F\int dx  \left[-\hat  \psi^\dagger_+ i\partial_x \hat \psi _+ 
+\hat \psi^\dagger_- i\partial_x\hat  \psi _-
\right] .
\end{align}

The continuum fermion field is $\hat \Psi(x) \equiv %\lim_{{a}\to0} \frac1{\sqrt{{a}}} \hat c_r \approx 
e^{i k_F x}\hat \psi_+(x)+e^{-i k_F x}\hat \psi_-(x)$, where $x %= r{a} 
\in [0,\ell]$ is the position along the GW. Defining an unfolded fermion field $\hat \psi(x) \equiv \hat \psi_+(x)$ and $\hat \psi(-x) \equiv \hat \psi_-(x)$ over $x\in[-\ell,\ell]$, we write the effective GW Hamiltonian (without disorder and interactions)
%%%%%
\begin{align}
\hat H_{f}^{\text{eff}}=-iv_F\int_{-\ell}^\ell dx\, \hat\psi^\dagger(x)\partial_x\hat \psi(x).
\end{align}
For the Kitaev case we switch to a Majorana basis by writing the fermion field in the GW as $\hat \psi(x,t)=\frac{1}{\sqrt{2}}\left[\hat \eta_1(x,t)+i\hat \eta_2(x,t)\right]$, where $\hat \eta_j=\hat\eta_j^\dagger$, $j=1,2$. 
The boundary condition on this fermion field is $\hat\psi(\ell) = \zeta \hat\psi(-\ell)$, where the sign $\zeta = e^{2ik_F \ell} = +1$ $(-1)$ corresponds to a GW with (without) an extended resonant state at the center of the band.

In its topological phase, the gapped systems admit midgap TBSs that prior to the quench are localized at the ends of the system. For the SSH model, these are fermion bound states, whose annihilation and creation operators we shall denote by $\hat f_a$ and $\hat f_a^\dagger$. For the Kitaev model, these are Majorana bound states which we shall denote by $\hat \gamma_a$. Here $a=L$ ($R$) denotes the TBS interfaced with the GW to the left (right) at position $x_{L}=0$ ($x_{R}=\ell$). In the low-energy theory, it is sufficient to consider only the coupling of the TBSs to the GW,
%%%%%
\begin{align}
\hat H_{\text{t}}^{\text{eff}}= 2 \sum_a\lambda_a(t) \hat\psi(x_a) \hat \alpha _a^\dagger+\text{h.c.},
\end{align}
%%%%%
where $\hat \alpha_a=\hat f_a$ ($\hat \alpha_a=i\hat \gamma_a/\sqrt{2}$) for SSH (Kitaev) models, respectively, and $x_L=0$, $x_R=\ell$. 
The factor of $2$ accounts for the fact that the full fermion field $\hat\Psi(x_a) = 2\hat\psi(x_a)$, up to a phase $e^{ik_F x_a}$ that is absorbed into $\lambda_a$.
Note that %in the continuum limit, $\lambda_e = \lim_{a\to0}\sqrt{a}\lambda_{rs}$, where $\lambda_{rs}$ is the tunneling energy in the lattice Hamiltonian; thus, 
$\lambda_a$ in the continuum Hamiltonian has dimensions of $\sqrt{\text{velocity}\times\text{energy}}$ (in natural units). The coupling to the end Majorana fermions is
\begin{eqnarray}
\label{eq:maj-coupling} 2i\sum_{a, j}\lambda_{a j}(t)\hat \eta_j(x_a)\hat \gamma_{a},
\end{eqnarray}
with $\lambda_{a i}(t)= \lambda_{a}(t)\sin[(\phi_a+\Lambda_a+\pi(i-1))/2]$. Here $\phi_a$ is the phase of the $a$ superconductor and $\Lambda_L=0$ ($\Lambda_R=\pi$) is associated with real (imaginary) Majorana modes. 

\subsection{Derivation of the TBS dynamics}
In the following we assume $\lambda_a\in\mathbb{R}$ and start with the case that $\lambda_L\neq 0$, $\lambda_R=0$. For a single Majorana TBS coupled to the GW only $\hat\eta_2$ remains coupled, thus we obtain similar equations of motion for the Kitaev and SSH case. We unify the notation by setting $\hat \nu=\hat \psi$ ($\hat \nu=\hat \eta/\sqrt{2}$, $\hat \eta\equiv \hat \eta_2$) for SSH (Kitaev). Thus we obtain the equations of motion Eq.~\eqref{eq:motion}. Integrating the first equation around $x=0$, we obtain $ v_F\hat \nu(0^+,t)= v_F\hat \nu(0^-,t)+i 2\lambda_L \hat \alpha_L(t)$. Resolving the discontinuity of the lead mode at $x=0$ by writing $\hat \nu(0,t)=[\hat \nu(0^+,t)+\hat \nu(0^-,t)]/2$, the second equation yields, $(\partial_t+\Gamma)\hat \alpha_L(t) = 2i \lambda_L\hat \nu(0^-,t)$, where $\Gamma=2|\lambda_L|^2/v_F$.
 This equation can be solved as
%%%%%
\begin{equation}
\hat \alpha_L(t)= e^{-\Gamma t} \hat \alpha_L(0) +2ie^{-\Gamma_L t}\int_0^t \lambda_L(s)e^{\Gamma s} \hat \nu(0^-,s)\, ds.
\end{equation}
%%%%%
In this linear model $\hat \nu(x,t)=\hat \nu(x-v_Fs,t-s)$ for $(x-v_Fs)x>0$, i.e. as long as the tunneling site at $x=0$ is not crossed. Thus, at $x=0^-$ and for $0<t<\tau_r$, we have a free field $\sum_\omega e^{-i\omega t}\hat \nu_0(\omega) \equiv \hat \nu_0(t)$. Complemented by the boundary condition $\hat \nu(x+v_F\tau_r,t)=\zeta\hat \nu(x,t)$, we have
%%%%%
\begin{align}
\hat \nu(0^-,s) &= \zeta\hat \nu(2\ell^+,s) \nonumber\\
&
= \zeta\hat \nu(0^+,s-\tau_r) \nonumber\\
&= \zeta\hat \nu(0^-,s-\tau_r) + i  (2\lambda_L/v_F) \zeta \Theta(s-\tau_r) \hat \alpha_L(s-\tau_r) \nonumber \\
& = \cdots \nonumber\\
&= \zeta^{n}\hat \nu(0^-,s-n\tau_r) \nonumber\\& 
+ i(2\lambda_L/v_F)\sum_{j=1}^n \zeta^j \Theta(s-j\tau_r)\hat \alpha_L(s-j\tau_r),
\end{align}
%%%%%
where $n$ is an integer. Choosing $n=\lfloor t/\tau_r\rfloor$ we find a recursive relation that we can iterate to find Eq.~(\ref{eq:f}) of the main text. %Thus, for $n=0$ we get $g_L(t)=e^{-\Gamma t}$. 
This can also be written as
%%%%%
\begin{equation}
g_L(t) = \sum_{j=0}^{n}  \zeta^je^{-\Gamma(t-j\tau_r)}L_j^{(-1)}\big(2\Gamma(t-j\tau_r)\big),
\end{equation}
%%%%%
where $L_j^{(-1)}$ is a generalized Laguerre polynomial.
For $\Gamma\tau_r\gg1$, the terms with $j<n$ contribute negligibly and $g_L(n\tau_r + s) \approx \zeta^n \varphi_n(\Gamma s)$, where $\varphi_n(x) = e^{-x}L^{(-1)}_n\left(2 x\right)$.
%f_1(t;\tau_r) &\approx - 2 \zeta \Gamma (t-\tau_r)e^{-\Gamma(t-\tau_r)}, \\
%f_2(t;\tau_r) &\approx -2\Gamma (t-2\tau_r)[1-\Gamma (t-2\tau_r)]e^{-\Gamma(t-2\tau_r)}.
%\varphi_n(x) = e^{-x}\sum_{k=1}^{n} 
%\multiset{k}{n-k} 
%\binom{n-1}{k-1}
%\frac{(-2x)^k}{k!}

To find the asymptotic expression for large $n$ when $\Gamma\tau_r\gg1$, we note that $\varphi_n$ satisfies a one-dimensional Schr\"{o}dinger equation with potential $-2 n/x$, mass $1/2$ and energy $-1$:
%%%%%
\begin{eqnarray}
	-\partial_x^2 \varphi_n+\left(1 -\frac{2 n}{x}\right) \varphi_n=0.
\end{eqnarray}
%%%%%
We employ a semi-classical approach to analyze the behavior of the largest (last) peak, obtained near the turning point, $x_{0}=2 n$. In this region we linearize the potential $x=x_0+\tilde{x}$
%\begin{eqnarray}
%	-\partial_{\tilde{x}}^2 \varphi_n+\left(1-\frac{2 n}{x_0}+\frac{2n}{x_0^2}\tilde{x}\right) \varphi_n=0.
%\end{eqnarray}
to get the Airy equation whose solution gives the following approximate behavior near $x_0$
%\begin{align}
%\tilde{y}=\Upsilon(n) \text{Ai}\left(\dfrac{\tilde{x}}{2 %(2n)^{1/3}}\right)
%\end{align}
\begin{align}
\varphi_n(\tilde{x})\approx \Upsilon(n) \text{Ai}\left(\dfrac{ \tilde{x}}{ (2n)^{1/3}}\right),
\end{align}
where $\text{Ai}$ is the Airy function and $\Upsilon(n)$ is a normalization constant. The global extrema of $\text{Ai}$ is the last oscillating peak value before the decay near $x_0$ and hence occurs close to the origin. We evaluate it to the third order around $\tilde x=0$ and obtain 
$
\text{Ai}^{\text{max}}
%\approx \dfrac{\Gamma\left(\frac{1}{3}\right)}{2 \times {3}^{1/6} \pi }+\dfrac{4 \sqrt{\pi }}{3^{17/12} \Gamma \left(\frac{1}{3}\right)^2}
\approx 0.5634.
$
The normalization constant 
%is given by 
%\begin{align}
%\Upsilon(n)=
%\left[2 n \text{Ai}\left(-(2 n)^{2/3}\right)^2+(2 n)^{1/3}\text{Ai}'\left(-(2 n)^{2/3}\right)^2\right]^{-\frac{1}{2}},
%\end{align}
can be evaluated exactly and has the asymptotic behavior
$
\Upsilon(n\gg 1 )=\frac{\sqrt{\pi }}{(2n)^{1/3}}.
$
Altogether we find the asymptotic approximation
%%%%%
\begin{equation}
|\varphi^{\mathrm{max}}_{n\gg 1}| \approx  \frac{0.79256 }{n^{1/3}},
 \end{equation}
 %%%%%x
quoted in the main text.

%The asymptotic behavior for large $n\gg1$ is dominated by $k=n$ and $\varphi_n(s) = e^{-\Gamma s}(-\Gamma s)^n/n!$. The maximum of this function is obtained at $\Gamma s = n$. Using the the asymptotic behavior of the factorial (Stirling's formula) $n! \approx n^n e^{-n}/\sqrt{2\pi n}$, we have the asymptotic expression,
%%%%%
%\begin{equation}
%\varphi_{n\gg1}^\text{max} \approx \frac{(-1)^n}{\sqrt{2\pi n}}.
%\end{equation}
%%%%%%

When two TBSs are coupled to $\hat\eta$ at $x_L=0$ and $x_R=\ell$, repeating the iterative procedure outlined in the previous section we obtain the following integral equations

\begin{eqnarray}
\nonumber \hat\alpha_R(t)&=&\mathcal{Y}_R[\hat\nu_0]+e^{-\Gamma t}\hat\alpha_R(0)-2\Gamma e^{-\Gamma t}\int_0^t ds\, e^{\Gamma s}\\
\nonumber &&\times\left[\sum_{j=1}^{\lfloor \frac{s}{\tau_r}+\frac{1}{2} \rfloor}\zeta^j\theta(s+\frac{\tau_r}{2}-j\tau_r)\hat\alpha_L(s+\frac{\tau_r}{2}-j\tau_r)\right.\\
&&\quad \left.+\sum_{j=1}^{\lfloor \frac{s}{\tau_r} \rfloor}\zeta^j\theta(s-j\tau_r)\hat\alpha_R(s-j\tau_r)\right],
\end{eqnarray}
\begin{eqnarray}
\nonumber \hat\alpha_L(t)&=&\mathcal{Y}_L[\hat\nu_0]+e^{-\Gamma t}\hat\alpha_L(0)-2\Gamma e^{-\Gamma t}\int_0^t ds\, e^{\Gamma s}\\
\nonumber &&\times\left[\sum_{j=1}^{\lfloor\frac{s}{\tau_r}+\frac{1}{2} \rfloor}\zeta^{j+1}\theta(s+\frac{\tau_r}{2}-j\tau_r)\hat\alpha_R(s+\frac{\tau_r}{2}-j\tau_r)\right.\\
&&\quad \left.+\sum_{j=1}^{\lfloor \frac{s}{\tau_r} \rfloor}\zeta^{j}\theta(s-j\tau_r)\hat\alpha_L(s-j\tau_r)\right],
\end{eqnarray}
which we can solve by iterations to obtain Eqs.~(\ref{eq:g-1}),(\ref{eq:g-2}) of the main text.

%%%  Effective theory of fractional charge propagation  %%%
%\vspace{-2mm}
\subsection{Effective theory of fractional charge propagation}

The occupation of the bound state $N(t)\equiv\langle \hat f^\dagger(t) \hat f(t) \rangle = e^{-2\Gamma t} N(0)$ decays for $0<t<\tau_r$, and is revived for $\tau_r<t<2\tau_r$ as $N(t) \approx 4\Gamma^2 (t-\tau_r)^2e^{-2\Gamma(t-\tau_r)}$ with a maximum value $N(\tau_r+1/\Gamma)/N(0)=4/e^2\approx0.54$, irrespective of $\zeta$.
%%%%%
%where the free field form
%%%%%%
%\begin{align}
%\tilde\psi_{0k}(t;\tau) = f_k(t;\tau) [\partial_t\tilde\psi_0(t)] + [(\partial_t + \Gamma)f_k(t;\tau)] \tilde\psi_0(t).
%\end{align}
%%%%%%

For the lead, we have %\babak{I don't understand the 3rd line and later below: What happened to $s$? What is $\tilde\psi_{0k}$?}
%%%%%
\begin{align}
\hat \psi(x,t) &= \hat \psi(0^+,t-x/v_F) =- \frac{i}{2\lambda}(\partial_s-\Gamma)\hat f(s) \big\vert_{s=t-x/v_F}.
\end{align}
%%%%%
We can calculate the density in terms of the unfolded fermion field, %\babak{Is this for $0<t<\tau$?}
%%%%%
\begin{align}
\rho(x,t)  
&= \langle \hat \Psi^\dagger (x,t)\hat \Psi (x,t)  \rangle \nonumber \\
&= \langle\hat \psi^{\dagger}(x,t)\hat \psi(x,t)\rangle+\langle\hat \psi^{\dagger}(-x,t)\hat \psi(-x,t)\rangle \nonumber \\
&+ e^{-2ik_{F}x}\langle\hat \psi^{\dagger}(x,t)\hat \psi(-x,t)\rangle \nonumber \\
&+ e^{2ik_{F}x}\langle\hat \psi^{\dagger}(-x,t)\hat \psi(x,t)\rangle.
\end{align}
%%%%%
%which we write in terms of the unfolded field $\psi(-x)=\psi_L(x)$ and $\psi(x)=\psi_R(x)$ and get
For example, in a a semi-infinite chain $x>0$, we find
\begin{align}
\nonumber \rho(x,t)=&\frac2\xi \theta(x - v_F t)e^{-2(x-v_Ft)/\xi}\left[ N(0)-\frac{1}{2}\right] \\ \nonumber
&+\frac1\pi \sin(2 k_F x)\left\{\frac{1}{2x}- \frac2\xi \theta(x-v_Ft)e^{-2 x/\xi} \right.\\ & \left.
\left[\mathrm{E}_1\left(-\frac{2x}{\xi}\right)-\mathrm{E}_1\left(-\frac{x-v_Ft}{\xi}\right)\right]\right\},
\end{align}
where $\xi = v_F/\Gamma$ is the leakage length, and E$_1(x)=\int_{x}^\infty \frac{e^{-t}}{t}dt$ is the exponential integral. At half-filling, the Friedel oscillations from the second term vanish and the only contribution comes from the exponentially localized charge from the first term contributed by the propagating soliton. In a finite geometry, this propagating charge is reflected and returns to the original site of the soliton. The magnitude of the charge on the original site is given by the square of the envelop function  $|g_L(t)|^2$.

\subsection{Boundary changing operators and monodromies}

The non-interacting, clean limit of the gapless wire is described
by two copies of the chiral part of the $\mathbb{Z}_{2}$ Ising conformal
field theory. Each copy contains three primary fields, the identity
$I$ (with conformal dimension $h_{I}=0$), a real fermion field $\eta(x,t)$
($h_{\eta}=1/2$) and the twist field $\sigma(x,t)$ ($h_{\sigma}=1/16$).
Here $\sigma$ acts as the BCO generating the
quench in the limit $\lambda\gg v_{F}/\sqrt{\ell}$. Therefore,
the state following the quench with a single topological wire is $|\Omega^{(1)}(t)\rangle=\sigma(0,t)|\Omega_0\rangle$
(where $|\Omega_0\rangle$ is the state of the wire prior to the
time of the quench at $t_0=0$), and $|\Omega^{(2)}(t)\rangle=\sigma(\ell,t)\sigma(0,t)|\Omega_0\rangle$
for a quench with two topological wires. We consider the overlap between
the state at time $t$ and the state at time $0$, $\langle\Omega^{(j)}(t)|\Omega^{(j)}(0)\rangle$,
for $j$ topological wires ($j=1,2$). The fidelity for one and two wires is given by
\begin{equation}
\mathcal{F}^{(1)}(t)=|\langle\Omega_0|\sigma(0,t)\sigma(0,0)|\Omega_0\rangle|^{2},
\end{equation}
and
\begin{align}
\mathcal{F}^{(2)}(t)=|\langle\Omega_0|\sigma(0,t)\sigma(\ell,t)\sigma(\ell,0)\sigma(0,0)|\Omega_0\rangle|^{2}.
\end{align}
In the following we will therefore focus on calculating the correlators for the BCOs. 

In terms of coordinates $z_{i}$ spanning the complex plane, the correlator of two BCOs is
\begin{equation}
\langle\sigma(z_{1})\sigma(z_{2})\rangle=\frac{1}{z_{12}^{2h_{\sigma}}},
\end{equation}
where $z_{ij}\equiv z_{i}-z_{j}$. For four BCOs,
the correlation function is
\begin{equation}
\langle\sigma(z_{1})\sigma(z_{2})\sigma(z_{3})\sigma(z_{4})\rangle=\left(\frac{z_{13}z_{24}}{z_{12}z_{23}z_{14}z_{34}}\right)^{2h_{\sigma}}R(y),
\end{equation}
where $y=\frac{z_{12}z_{34}}{z_{13}z_{24}}$ and $R(y)$ is given
by
\begin{equation}\label{eq:R}
R(y)=\alpha_{+}R_+(y)+\alpha_{-}R_-(y),
\end{equation}
where $\alpha_{\pm}$ will be decided by the initial conditions and $R_\pm=\sqrt{1\pm \sqrt{1-y}}$. 

Our theory is defined on a strip $0<x<2\ell$, therefore we use the conformal transformation $z=\exp\left(\pi w/\ell\right)$ (and its inverse $w=\frac{\ell}{\pi}\ln z$) and the transformation rule for correlators
\begin{align}
\left\langle\prod_i\sigma(w_{i})\right\rangle=\left[\prod_{i=1}\left(\frac{dw}{dz}\right)_{w=w_{i}}^{-h_{\sigma}}\right]\left\langle\prod_i\sigma(z_{i})\right\rangle
\end{align}
to get for the correlator of two BCOs
\begin{align}
\langle\sigma(w_{1})\sigma(w_{2})\rangle=\left(\frac{\pi}{2\ell}\right)^{2h_\sigma}\left[\frac{1}{s(w_{12})}\right]^{2h_{\sigma}},
\end{align}
where $s(w)=\sinh\left(\frac{\pi w}{2\ell}\right)$, $w_{ij}=\mathsf{a}+i(x_{ij}-v_{F}t_{ij})$ and $\mathsf{a}$ is a short distance cutoff. For four BCOs we obtain
\begin{align}
& \langle\sigma(w_{1})\sigma(w_{2})\sigma(w_{3})\sigma(w_{4})\rangle\nonumber \\
& =\left(\frac{\pi}{2\ell}\right)^{4h_{\sigma}}\left[\frac{s(w_{13})s(w_{24})}{s(w_{12})s(w_{23})s(w_{14})s(w_{34})}\right]^{2h_{\sigma}}R\left(\mathsf{y}\right),
\end{align}
where $R(\mathsf{y})$ is given in Eq.~\eqref{eq:R} and $\mathsf{y}=\frac{s(w_{12})s(w_{34})}{s(w_{13})s(w_{24})}$.

%\begin{figure}
%	%\includegraphics[width=0.49\textwidth]{Daniel/Plots/fig9_interactions/fig9.pdf}
%	\includegraphics[width=3.4in]{fig10.pdf}
%	\caption{Fidelities and relative statistical phase corresponding to the numerical results presented in Fig.~\ref{fig:FidelityAndPhase} of the main text, but as extracted from the correlators of BCOs.}   
%	\label{fig:BCO_SSH} 
%\end{figure}

For the Kitaev case, the relevant conformal blocks are $B_\pm(\mathsf{y})=[R_+(\mathsf{y}) \pm(-1)^{\lfloor \frac{t}{\tau_r}+\frac{1}{2}\rfloor} R_-(\mathsf{y})]/2$. For the SSH case, the correlation function is effectively squared. Care should be taken in squaring the conformal blocks. It turns out that the relevant conformal blocks are $D_\pm=B_+^2 \pm B_-^2$, for the case that the two TBS are occupied ($D_-$) and for the case that one is occupied and the other not ($D_+$). %The result for the fidelities associated with $D_\pm$ and their relative phase is presented in Fig.~\ref{fig:BCO_SSH}\,, which demonstrate remarkable agreement with the numerical results for the quench dynamics presented in Fig.~\ref{fig:FidelityAndPhase}\,.

%%%%%%%%%%%%%%%%%%%%%%%%%%%%%%%%%%%%%%
%%%  The trace formula and counting statistics  %%%
%%%%%%%%%%%%%%%%%%%%%%%%%%%%%%%%%%%%%%
\section{Trace Formula and Counting Statistics}\label{app:trace-formula}
In this Appendix, we derive the formulea for the entanglement entropy, the full counting statistics and the fermion parity. We also give expressions for the exact overlap of two superconducting states and obtain the Onishi formula which is used to calculate the overlap of states and extract the fidelity and phase.

For two bilinear forms $\hat{A}=\frac{1}{2}\hat\Psi^\dagger \mathcal{A} \hat\Psi$ and $\hat{B}=\frac{1}{2}\hat\Psi^\dagger \mathcal{B} \hat\Psi$, the following \emph{trace formulae} are useful for calculating expectation values:
%%%%%
\begin{align}
\tr~e^{\hat{A}} &= \sqrt{\det\left( \id+ e^{\mathcal{A}} \right)}\label{eq:trace1}, \\ %\left[\det\left(2\cosh\frac{\mathcal{A}}{2}\right)\right]^{\frac12},
\tr~e^{\hat{A}}e^{\hat{B}} &= \sqrt{\det\left( \id+ e^{\mathcal{A}}e^{\mathcal{B}}\right)}. \label{eq:trace2} %\left[\det\left(2\cosh\frac{\mathcal{C}}{2}\right)\right]^{\frac12}.
\end{align}
%%%%%%
Here, matrices $\mathcal{A}$ and $\mathcal{B}$, are taken, without loss of generality, to be particle-hole symmetric: $\mathcal{A} = -\Sigma_x \mathcal{A}^\trans \Sigma_x$ and similarly for $\mathcal{B}$, where $\Sigma_x$ is the Pauli matrix acting on the particle-hole basis in the Nambu space. Consequently, $\tr\mathcal{A}=0$, and $\det(\id+e^{\mathcal{A}}) = \det(2\cosh\frac{\mathcal{A}}2) = \det(\id+e^{-\mathcal{A}})$. %The matrix $\mathcal{C}$ is defined by $e^\mathcal{C}=e^\mathcal{A} e^\mathcal{B}$ and is also traceless and particle-hole symmetric. 
The sign of the square roots need to be determined by analytical continuity in the matrix elements. 

%%%  Entanglement entropy  %%%
\subsection{Entanglement Entropy}\label{app:EE}
We use the first trace formula to calculate the entanglement entropy of a subsystem $\mathsf{s}$ with reduced matrix  $\rho_\mathsf{s}$ using the replica trick,
%%%%%
\begin{equation}
S = \tr_\mathsf{s}(\hat\rho_\mathsf{s}\log\hat\rho_\mathsf{s}) = -\lim_{n\to 1}\partial_n \tr_\mathsf{s}\,\hat\rho_\mathsf{s}^n
\end{equation}
%%%%%
Starting with the ground state $\ket{\Omega}$ of the system, we write the reduced density matrix
%%%%%
\begin{equation}
\hat \rho_\mathsf{s} = \tr_{\bar{\mathsf{s}}}|\Omega\rangle\langle\Omega| = \kappa_\mathsf{s} e^{-\hat A_\mathsf{s}},\label{eq:reduced-density}
\end{equation}
%%%%%
where $\bar{\mathsf{s}}$ is the complementary subsystem to $\mathsf{s}$, $\hat A_\mathsf{s}$ is an operator involving only the degrees of freedom in $S$ and $\kappa_\mathsf{s}$ is a normalization constant such that $\tr_\mathsf{s}\,\hat\rho_\mathsf{s} = 1$.

For the ground state of a quadratic Hamiltonian one can show, using Wick's theorem, that $\hat A_\mathsf{s}$ is also quadratic and given by a Hermitian matrix $\mathcal{A}_\mathsf{s}$. Then,
%%%%%
\begin{align}
\tr\,\hat\rho^n_\mathsf{s}
	&= \kappa^n_\mathsf{s} \left[ \det\left(2\cosh \frac{n\mathcal{A_\mathsf{s}}}{2}\right)\right]^{\frac12}.
\end{align}
%%%%%
The constant is found by setting $n=1$ to be
%%%%%
\begin{equation}
\kappa_\mathsf{s} = \left[ \det\left(2\cosh \frac{\mathcal{A}_\mathsf{s}}{2}\right)\right]^{-\frac12} = \frac1{\sqrt{\det(\id+e^{\mathcal{A}_\mathsf{s}})}}.
\end{equation}
%%%%%
Replacing this in the previous equation and taking the derivative, using 
$$
\frac{d}{dz} \det \mathcal{M}=\det(\mathcal{M}) \tr\left(\mathcal{M}^{-1}\frac{d}{dz}\mathcal{M}\right),
$$
and $\log\det\mathcal{M}=\tr\log\mathcal{M}$ for a general matrix $\mathcal{M}$, and finally taking the limit $n\to 1$, we arrive at 
%%%%%
\begin{equation}\label{eq:ent-A}
S_\mathsf{s} = \frac{1}{2} \tr\log\left(2\cosh\frac{\mathcal{A}_\mathsf{s}}{2}\right)-\frac{1}{2} \tr\left(\frac{\mathcal{A}_\mathsf{s}}{2}\tanh\frac{\mathcal{A}_\mathsf{s}}{2}\right). 
\end{equation}
%%%%%%

The above expression can be further simplified by the relation
%%%%%
\begin{equation}\label{eq:AG}
\mathcal{A}_\mathsf{s}  =  \log \mathcal{G}_\mathsf{s}-\log(\id-\mathcal{G}_\mathsf{s}),
\end{equation}
%%%%%
where $\mathcal{G}$ is the Nambu Green's function (projected to $\mathsf{s}$)
%%%%%
\begin{equation}
\mathcal{G} \equiv \langle\Omega| \hat\Psi\nodag \hat\Psi^\dagger |\Omega\rangle =  \begin{pmatrix} G & F\\ F^\dagger & \id -G^\trans \end{pmatrix}, \label{eq:projected-correlator}
\end{equation}
%%%%%
with the normal and anomalous correlators, respectively,
%%%%%
\begin{align}
 G &=\langle\Omega|\: \hat c\: \hat c^{\dagger} |\Omega\rangle = G^\dagger, \\
 F &=\langle\Omega|\: \hat c \: \hat c^\trans |\Omega\rangle = -F^\trans.
\end{align}
%%%%%
%We defer the proof of this relation to Appendix~\ref{app:EE}. 
Note that $\id-\mathcal{G}=\Sigma_x \mathcal{G}^\trans\Sigma_x$. Using Eq.~(\ref{eq:AG}) in Eq.~(\ref{eq:ent-A}), we find
%%%%%
\begin{equation}\label{eq:ent-G}
S_\mathsf{s} = -\frac{1}{2} \tr \left[\mathcal{G}_\mathsf{s}\log \mathcal{G}_\mathsf{s}+(\id-\mathcal{G}_\mathsf{s})\log(\id-\mathcal{G}_\mathsf{s})\right].
\end{equation}

%%%%%%%%% FIG. 5 %%%%%%%%%%%
%%%%%%% PARITY - EXACT %%%%%%%%
%%%%%%%%%%%%%%%%%%%%%%%%
\begin{figure}
	\centering 
	\includegraphics[width=3.4in]{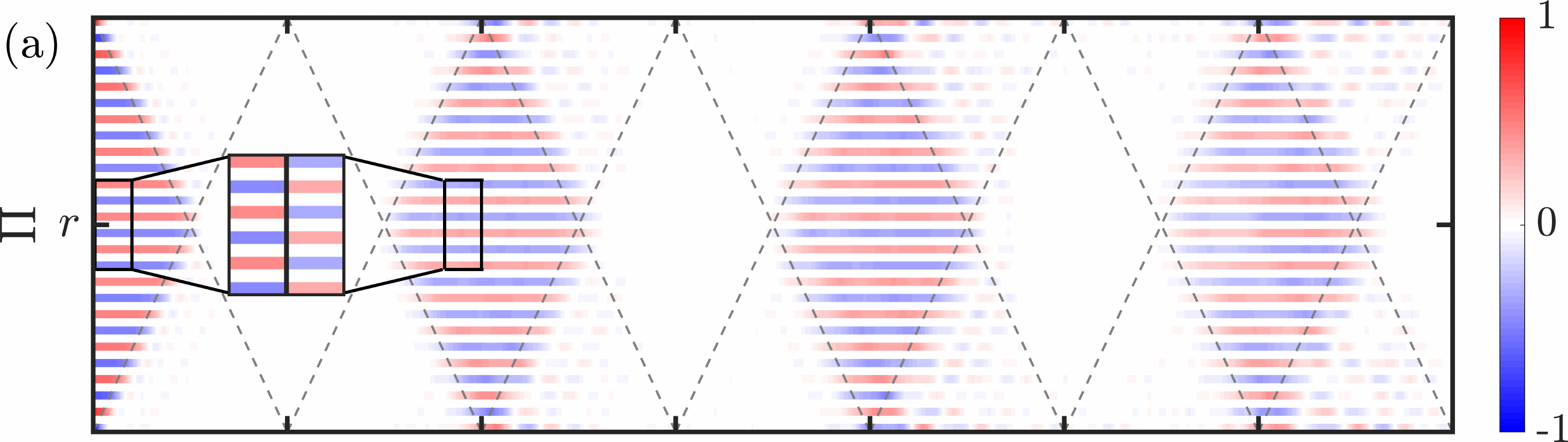}
	\hspace*{-0.05in}
	\vspace*{0.1in}
	\includegraphics[width=3.43in]{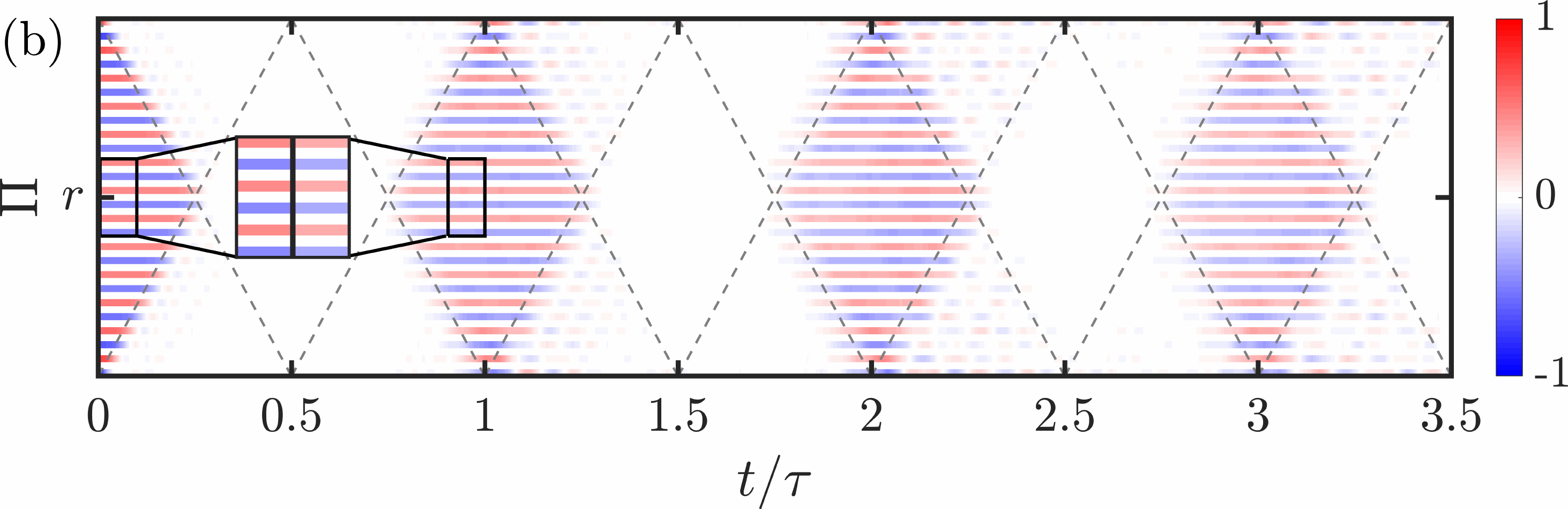}
	\vspace*{-0.3in}
	\caption{
		The fermion parity $\Pi$ to the left of a cut in the GW after a quench at $t_0 = 0$ couples ($\lambda_{L,R}= w$) the GW to two Kitaev chains with $\Delta_{L} =\Delta_R= 0.5w$  (top) and $\Delta_L=-\Delta_R=0.5w$ (bottom). The insets demonstrate the orthogonality (non-orthogonality) of the states for consecutive periods for $\Delta_L=\Delta_R$, top panel ($\Delta_L=-\Delta_R$, bottom panel).
	}
	\label{fig:parity-K-GW-K} 
\end{figure}
%%%%%%%%%%%%%%%%%%%%%%%%

%%%  Counting statistics and parity  %%%
\subsection{Counting statistics and parity}\label{app:CS}
We now employ the second trace formula to calculate the full counting statistics, generating the moments of the charge operator in subsystem $\mathsf{s}$. This generating function can be found from the expectation value of the operator
%%%%%
\begin{align}
	\hat{\chi}_\mathsf{s} (\phi)=e^{i\phi \hat{Q}_\mathsf{s}}=
	\exp\left[ \frac{i\phi}2 \left( \hat\Psi^\dag \Sigma_z \hat\Psi + N_\mathsf{s} \right)\right]
\end{align} 
%%%%%
where $N_\mathsf{s}$ is the number of sites in subsystem $\mathsf{s}$, and $\Sigma_{z}$ is the Pauli matrix in the Nambu space projected to $\mathsf{s}$. Using the trace formula, we obtain
%%%%
\begin{align}
\chi_\mathsf{s} (\phi) &\equiv %\langle \hat{\chi} \rangle =
	 \tr (\hat\rho_\mathsf{s} \hat\chi_\mathsf{s}) = \kappa_\mathsf{s} \tr\left(e^{-\hat{A}_\mathsf{s}}e^{i\phi\hat{Q}_\mathsf{s}}\right)  \nonumber\\
	&= e^{i\frac{\phi}{2} N_\mathsf{s}} \sqrt{\frac{\det\left(\id+e^{-\mathcal{A}_\mathsf{s}}e^{i\phi\Sigma_z}\right)}{\det{\left(\id+e^{\mathcal{A}_\mathsf{s}}\right)}}}.
\end{align}
%%%%%
The sign of the square root must be determined by analytic continuation. We note that by definition, the generating function is periodic,  $\chi_\mathsf{s}(\phi+2\pi)=\chi_\mathsf{s}(\phi)$. Since the prefactor $\exp(i\phi N_\mathsf{s}/2)$ is (anti-)periodic in $\phi$ for even (odd) $N_\mathsf{s}$, while the determinant under the square root is always periodic, we take the square root to have a (no) branch cut on the real axis of the complex variable $e^{i\phi}$ for even (odd) $N_\mathsf{s}$. This restores the overall periodicity of the generating function. 

Using Eq.~(\ref{eq:AG}) $\chi_\mathsf{s}$ can be written as
%%%%%
\begin{equation}
\chi_\mathsf{s}(\phi) = e^{i \frac{\phi}2N_\mathsf{s}}\sqrt{\det\left( \id-\mathcal{G}_\mathsf{s} + e^{i\phi\Sigma_z}\mathcal{G}_\mathsf{s}\right)}.
\end{equation}
%%%%%
We define $z=(1-e^{i\phi})^{-1}$ and, noting $\det\Sigma_x = (-1)^{N}$, we find up to a sign,
%%%%%
\begin{equation}
\chi_\mathsf{s}(z) = (2z)^{-N_\mathsf{s}}\sqrt{\det[(2z-1)i\Sigma_y+(\id - 2\mathcal{G}_\mathsf{s})\Sigma_x]}.
\end{equation}
%%%%%
Note that the determinant is now being taken of an anti-symmetric matrix. Using the fact that for an anti-symmetric matrix $\mathcal{O}$, $\det\mathcal{O} = \pf^2\mathcal{O}$, we have
%%%%%
\begin{align}
\chi_\mathsf{s}(z) = (2z)^{-N_\mathsf{s}}s_{N_\mathsf{s}}\pf\left[(2z-1)i\Sigma_y+(\id - 2\mathcal{G})\Sigma_x\right],
\end{align}
%%%%%
where the sign $s_{N_\mathsf{s}}$ is determined below. The fermion parity $\Pi_\mathsf{s}$ is found by setting $\phi=\pi$, i.e. $z=\frac12$; thus,
%%%%%
\begin{align}
\Pi_\mathsf{s} = s_{N_\mathsf{s}}\pf\left[ \Sigma_x(\id-2\mathcal{G}_\mathsf{s}) \right],
\end{align}
%%%%%
The prefactor sign here is determined by analytic continuation in elements of $\mathcal{G}_\mathsf{s}$, since these depend analytically on the ground state of the system. Specifically, we may take the limit where the state $|\Omega\rangle$ approaches the ground state of a normal system, whence $F\to0$, with the fermion parity becoming $\Pi^{(0)}_\mathsf{s}=\det(2G_\mathsf{s}-1)$; thus, in this limit
%%%%
\begin{align*}
\pf[\Sigma_x(\id-2\mathcal{G}_\mathsf{s})] 
&\to \pf\begin{pmatrix}0 & 2G^\trans_\mathsf{s} -1 \\ 1-2G_\mathsf{s} & 0 \end{pmatrix} \\
&= (-1)^{N_\mathsf{s}(N_\mathsf{s}-1)/2}\Pi^{(0)}_\mathsf{s},
\end{align*}
%%%% 
where we used the property of the pfaffian that $\pf\begin{psmallmatrix}0 & M \\ -M^T & 0\end{psmallmatrix}=(-1)^{n(n-1)/2}/2$ for an arbitrary matrix $M$ of size $n\times n$. Therefore,
%%%%%
\begin{equation}
s_{N_\mathsf{s}} = (-1)^{\frac12{N_\mathsf{s}(N_\mathsf{s}-1)}}.
\end{equation}
%%%%%
As an example for the applicability of this formula, the parity for a quench coupling two Kitaev chains to a GW \cite{dahan2017non} is plotted in Fig.~\ref{fig:parity-K-GW-K}\,.

\section{Robustness}% to disorder and interactions}
\label{app:robustness}
In this Appendix, we formulate and study of the effects of short-range interactions and disorder in the GW.

\subsection{Effects of short-range interactions}

Now we apply bosonization techniques for the GW and write the right and left-moving fields of Eq.~\eqref{Lut1} as
\begin{align}
\hat \psi_{\pm}(x)\sim \frac{1}{\sqrt{2\pi \mathsf{a}}}e^{-i\sqrt{2\pi} \hat \phi _{\pm}(x)},
\end{align}
where $\phi_\pm$ are bosonic fields. It can be shown that the corresponding normal ordered densities are related to the derivative of the bosonic fields
\begin{equation}
\hat \rho_{\pm}(x) \sim \mp \frac{1}{\sqrt{2\pi}} \partial_x \hat \phi _{\pm}(x),
\end{equation}
and that the free part of the Hamiltonian in Eq.~\eqref{Lut1} can be rewritten in terms of the densities as 
\begin{align}\label{free_Lut}
\hat {H}_f= \frac{v_F}{2}\int dx \left[
\left( \partial_x\hat \phi _{+}(x)\right)^2
+\left(\partial_x \hat \phi _{-}(x)\right)^2
\right].
\end{align}
The interacting part of Eq.~\eqref{eq:GW} can be rewritten as
\begin{align}\label{H_int}
&\hat H_{\text{int}}=u\int dx \left[\hat \rho_+ (x)\hat \rho_+ (x+1)+\hat \rho_{-} (x)\hat \rho_- (x+1) \right. \nonumber \\ \nonumber
& \;
+\hat \rho_+ (x)\hat \rho_- (x+1)+\hat \rho_+ (x)\hat \rho_- (x+1)\\ \nonumber
& \;
+e^{2ik_F}\hat \psi^\dagger _+(x)\hat \psi _-(x)\hat \psi^\dagger _-(x+1)\hat \psi _+(x+1) +\mathrm{h.c.}\\ \nonumber
& \;
\left.+e^{-2ik_F(2x+1)}\hat \psi^\dagger _+(x)\hat \psi _-(x)\hat \psi^\dagger _+(x+1)\hat \psi _-(x+1)+\mathrm{h.c.}\right].\\
\end{align}

%%%%%%%%%%%%%%%%%%%%%%%%
%%%%%%%%% FIG. 8 %%%%%%%%%%%
%%%%%%% DISORDER %%%%%%%%%%
%%%%%%%%%%%%%%%%%%%%%%%%
\begin{figure*}
	\centering 
	\includegraphics[width=0.49\textwidth]{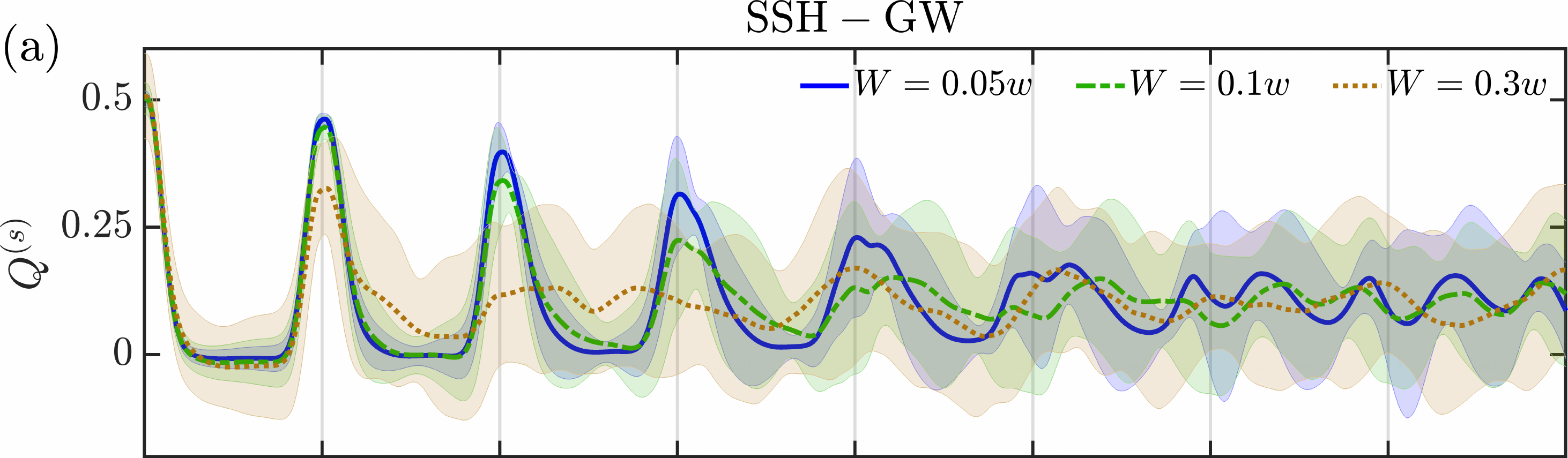}
	\includegraphics[width=0.49\textwidth]{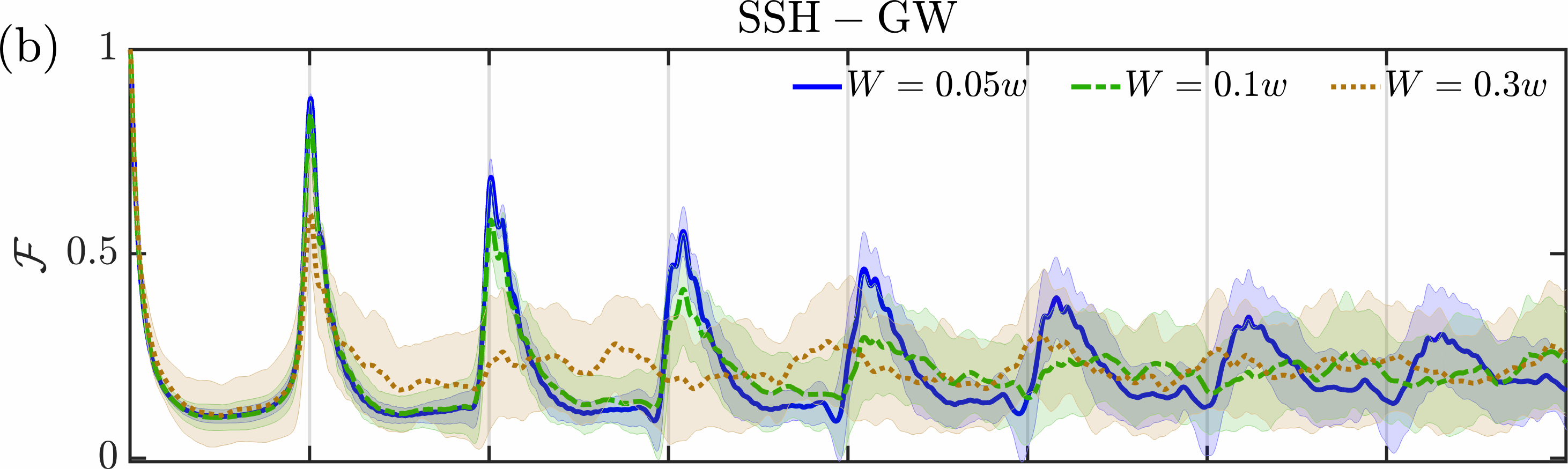}\\
	\includegraphics[width=0.49\textwidth]{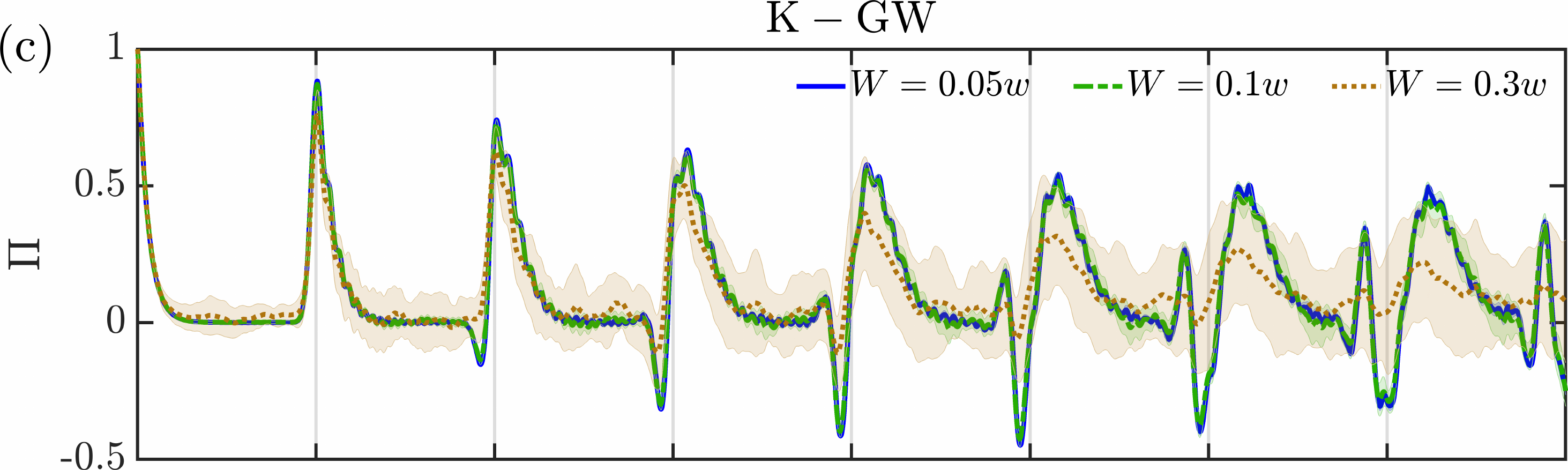}
	\includegraphics[width=0.49\textwidth]{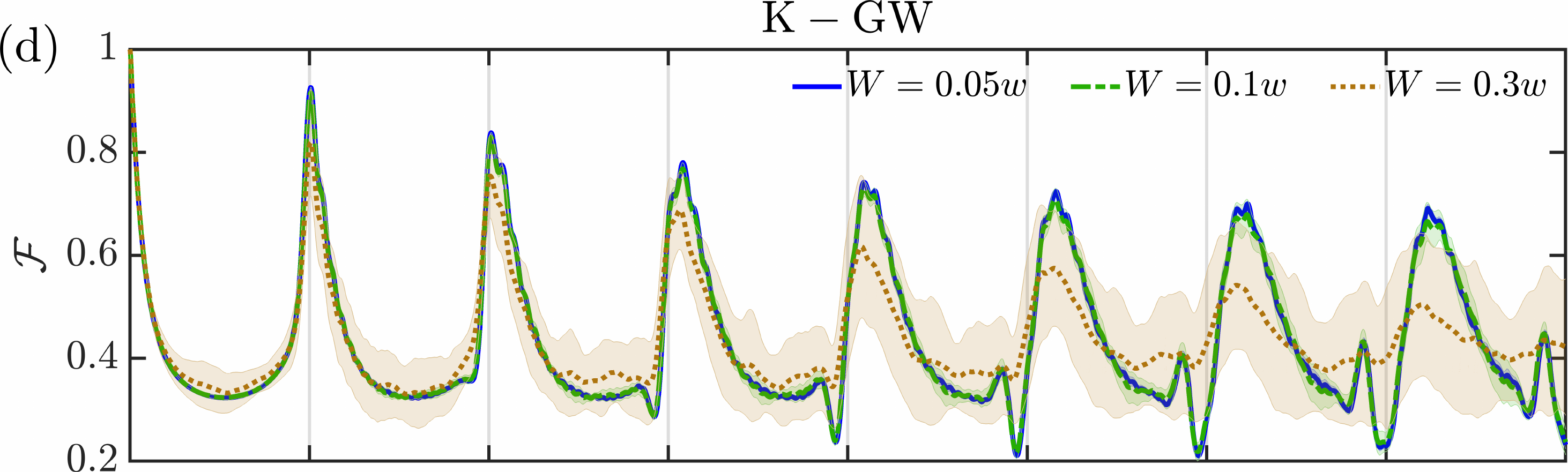}\\
	\includegraphics[width=0.49\textwidth]{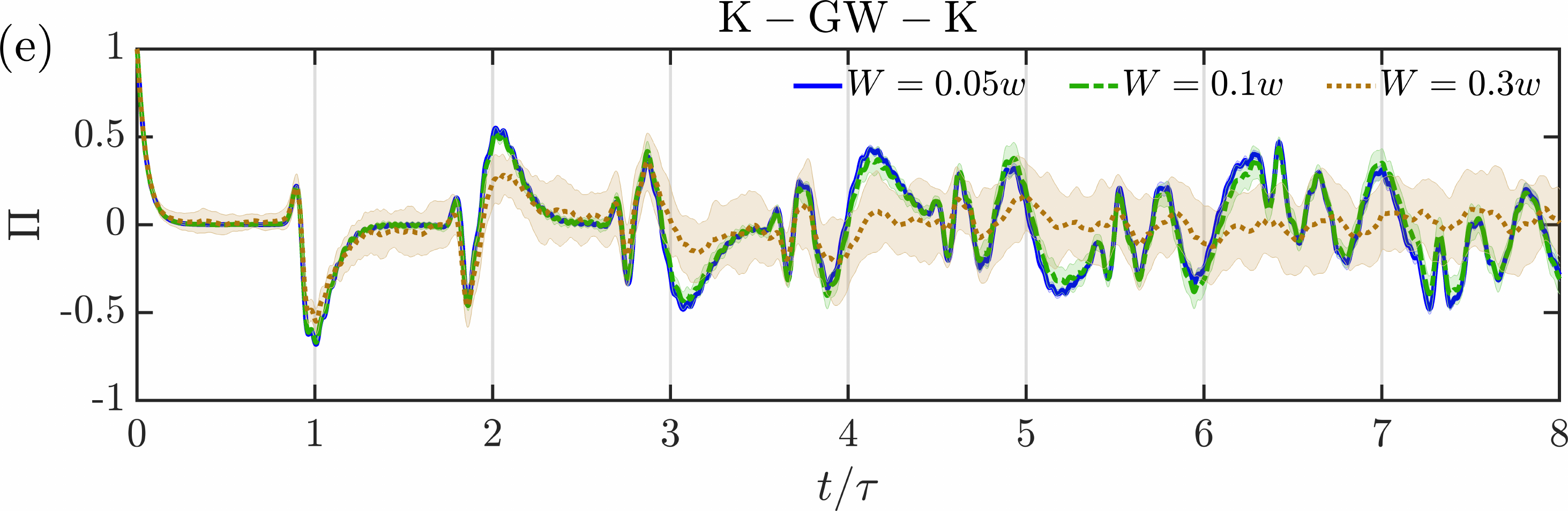}
	\includegraphics[width=0.49\textwidth]{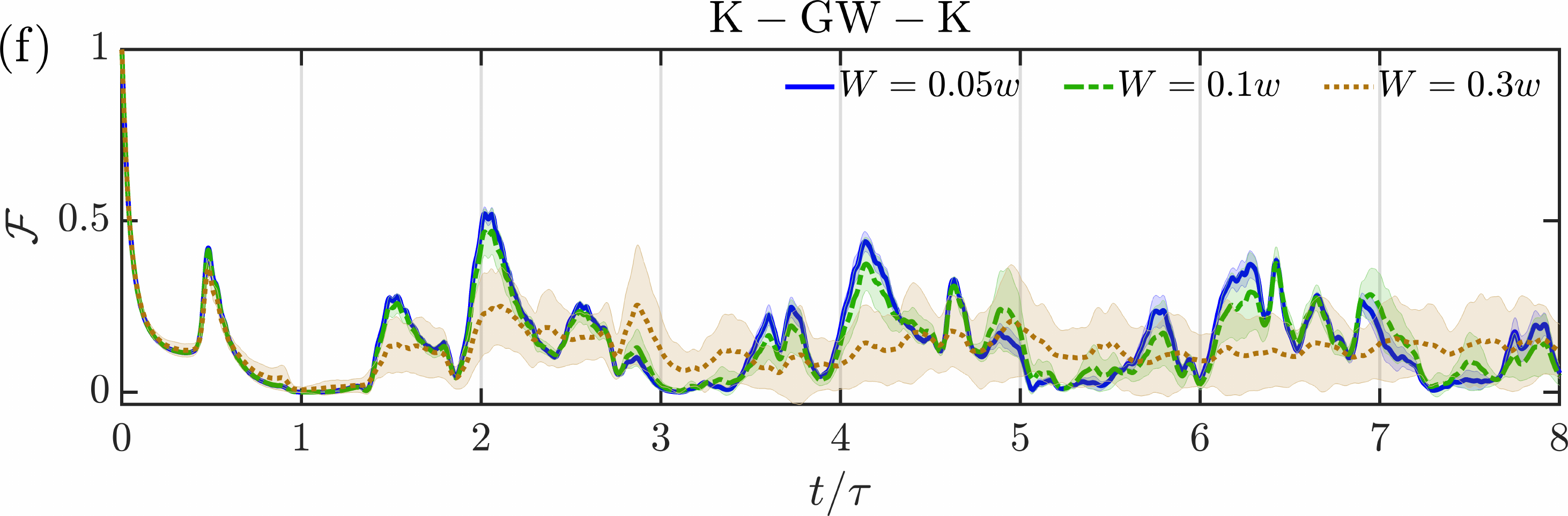}	
	\vspace{-0.2cm}
	\caption{Fidelity, parity and soliton charge in the presence of disorder in the GW.
		Panels (a) and (b): a sudden quench of a SSH chain ($N_L=50$, $m_L=0.8w$, $\lambda_L=0.4w$) to a GW ($N=100$). 
		Panels (c) and (d) [(e) and (f)]: a sudden quench of one (two) Kitaev chain (chains) with $N_L=20$ ($N_{L,R}=20$) and $\Delta_L=-0.7w$ ($\Delta_{L,R}=-0.7w$) and $\lambda_L=0.4w$ ($\lambda_{L,R}=0.4w$) to a GW with $N=100$. Panel (a) [(c) and (e)] shows the soliton charge (parity) behavior on the far edge of the SSH (Kitaev), wherein panels (b), (d) and (f) display the fidelities for the corresponding evolutions. The lines represent the mean value over $25$ realizations for a given disorder strength: $W=0.05w$ (dotted green line), $W=0.1w$ (dashed orange line) and $W=0.3w$ (blue line). Shaded areas represent the standard deviation in the disorder average.}
	\label{fig:disorder} 
\end{figure*}
%%%%%%%%%%%%%%%%%%%%%%%%

Dropping the oscillating fourth line, noting the relations
\begin{align}
&\hat \rho_{\pm} (x)\hat \rho_{\pm} (x+1)= \frac{1}{2\pi}\left(\partial_x \hat \phi _{\pm}\right)^2,\\
&\hat \rho_+ (x)\hat \rho_- (x+1)=-\frac{1}{2\pi}\partial_x \hat \phi _+\partial_x \hat \phi _-,\\
&\nonumber \hat \psi^\dagger _+(x)\hat \psi _-(x)\hat \psi^\dagger _-(x+1)\hat \psi _+(x+1)
\\&\approx 
-\frac{1}{(2\pi \mathsf{a})^2} \left[\pi \left(\partial_x\hat \phi _+(x)- \partial_x\hat \phi _-(x)\right)^2 \right],
\end{align}
and substituting everything into Eq.~\eqref{H_int}, we get
\begin{align}
\hat {H}_{\text{int}} %&=\frac{u}{2\pi}\int dx \left[\left(1-\cos(2k_F) \right) \left(\partial_x\hat \phi _+(x)- \partial_x\hat \phi _-(x)\right)^2  \right]\\ \nonumber
=\frac{u}{2\pi}\int dx \left[2\sin^2(k_F)\left(\partial_x\hat \phi _+(x)- \partial_x\hat \phi _-(x)\right)^2  \right],
\end{align}
where we have taken $\mathsf{a}$ as the lattice spacing (and set it equal to one), and dropped any constants. Adding to the free part of the Hamiltonian, Eq.~\eqref{free_Lut},
we obtain the Luttinger liquid (LL) Hamiltonian $H_{\text{LL}}=H_f+H_\text{int}$,
\begin{align}
\hat  H_{\text{LL}}= \frac{v_F}{2} \int dx \left\lbrace
\left(1+\frac{g_4}{2\pi v_F} \right)\left[\left(\partial_x \hat \phi_+ \right)^2+\left(\partial_x \hat \phi_- \right)^2\right]\right. \nonumber \\   \left.
-\frac{g_2}{\pi v_F}
\partial_x \hat \phi_+\partial_x \hat \phi_-
\right\rbrace,
\end{align}
where $g_2=g_4=4u \sin^2(k_F)$.

Introducing the bosonic field $\hat \phi$ and its dual field $\hat{\theta}$
\begin{align}
\hat{\phi}= \frac{\hat\phi_--\hat\phi_+}{\sqrt{2}},\quad \hat{\theta}=\frac{\hat\phi_-+\hat\phi_+}{\sqrt{2}},
\end{align} 
the Hamiltonian can be rewritten in the form 
\begin{equation}
\hat  H_{\text{LL}}=\frac{v}{2}\int dx \left[K\left(\partial_x\hat{\phi}\right)^2+\frac{1}{K}\left(\partial_x\hat{\theta}\right)^2 \right].
\end{equation}
Here we defined the Luttinger parameter
\begin{align}
K&=\sqrt{\frac{1+\frac{g_4}{2\pi v_F}-\frac{g_2}{2\pi v_F}}{1+\frac{g_4}{2\pi v_F}+\frac{g_2}{2\pi v_F}}}= \sqrt{\frac{1}{1+\frac{4u \sin^2(k_F)}{\pi v_F}}}
\nonumber \\ 
&\approx 1- \frac{2u \sin^2(k_F)}{\pi v_F}
=1-\frac{u \sin(k_F)}{\pi w} ,
\end{align}

and the Luttinger velocity,
\begin{align}
&v=v_F\sqrt{\left( 1+\frac{g_4}{2\pi v_F} \right)^2-\left( \frac{g_2}{2\pi v_F}\right)^2} \nonumber \\  
&=v_F\sqrt{1+\frac{4 u \sin^2(k_F)}{\pi v_F}}\approx v_F\left(1+ \frac{u \sin(k_F)}{\pi w}\right).
\end{align}
These approximations are valid for $|u/2w| \ll 1$. In the full range $-|2w|<u\leq |2w|$, $v$ and $K$ are obtained by the Bethe ansatz solutions quoted in the main text.

\subsection{Effects of disorder}

We study the effects of disorder on the quench by taking a finite width $W$ for the disorder potential, see Eq.~(\ref{eq:GW}). Technically, we take a large number of realizations and average the different observables at each time, while keeping track of the standard deviation. We take the disorder strength to be of the order of the level spacing (both below and above). The average values for the soliton charge (for a lead connected to SSH) and parity (for a lead connected with a Kitaev wire on one or two sides) are displayed in Fig.~\ref{fig:disorder}, as well as the fidelity (right panels). The standard deviation is displayed as shaded areas.

As evident from the graphs, for the Kitaev case the first few peaks remain remarkably sharp for disorder strengths that are as high as $30\%$ of the tunneling strengths. The SSH case is more sensitive to disorder, but the first peak remains quite sharp for comparable disorder strengths.

%We interpret these results as follows. For the SSH case the operators $\hat f$ and $\hat f^\dagger$ associated with the zero mode can couple locally to disorder via their associated density operator $\hat f^\dagger \hat f$. However, the Majorana fermions $\hat \gamma_L$ and $\hat \gamma_R$ can only form a local density like operator when they have overlapping support, which is the case only for a short period during the traversal compared with $\tau$. This suggests they are better protected against disorder.

%\bibliographystyle{prx}
%\bibliography{dynamics-TBS}

\end{document}